\documentclass[12pt]{iopart}
\usepackage{graphicx}
\begin{document}
%
%
\title [Frank-Kasper Compounds]{Magnetic, specific heat and electrical transport properties of Frank-Kasper cage compounds RTM$_2$Al$_{20}$ [R = Eu,Gd and La ; TM = V,Ti]}
\author{Ramesh Kumar K$^{1*}$, Harikrishnan S. Nair$^{1,2}$,  Reinke Christian$^{3}$, Thamizhavel A$^{4}$ and Andr\'{e} M. Strydom$^1$}
\address{$^{1}$ \scriptsize Highly Correlated Matter Research Group, Physics Department, P. O. Box 524, University of Johannesburg, Auckland Park 2006, South Africa.}
\address{\small$^{2}$ \scriptsize Department of Physics, Colorado State University, Fort Collins, CO 80523, USA}
\address{\small$^{3}$ \scriptsize Analytical Facility, P. O. Box 524, University of Johannesburg, Auckland Park 2006, South Africa.}
\address{\small$^{4}$ \scriptsize Department of Condensed Matter Physics and Material Sciences, Tata Institute of Fundamental Research, Mumbai, India}
\ead{kraamesh57@gmail.com, rkkamadurai@uj.ac.za}

\clearpage
\begin{abstract}
	Single crystals of Frank-Kasper compounds RTM$_2$Al$_{20}$ (R = Eu, Gd and La; TM = V and Ti) were grown by self-flux method and their physical properties were investigated through magnetization ($M$), magnetic susceptibility ($\chi$), specific heat ($C_P$) and electrical resistivity ($\rho$) measurements. Powder x-ray diffraction studies and structural analysis showed that these compounds crystallize in the cubic crystal structure with  the space group  $Fd\overline{3}m$. The magnetic susceptibility for the compounds EuTi$_2$Al$_{20}$ and GdTi$_2$Al$_{20}$ showed a sudden jump below the N{\'e}el temperature $T_N$ indicative of plausible double magnetic transition. Specific heat ($C_P$) and electrical resistivity ($\rho$) measurements also confirm the first-order magnetic transition (FOMT) and possible double magnetic transitions. Temperature variation of heat capacity showed a sharp phase transition and huge $C_P$ value for the (Eu/Gd)Ti$_2$Al$_{20}$ compounds Full width at half-maximum (FWHM) $<$ 0.2 K) which is reminiscent of a first-order phase transition and a unique attribute among RTM$_2$Al$_{20}$ compounds. In contrast, linear variation of $C_P$ is observed in the ordered state for (Eu/Gd)V$_2$Al$_{20}$ compounds suggesting a $\lambda$-type transition. We observed clear anomaly between heating and cooling cycle in temperature-time relaxation curve for the compounds GdTi$_2$Al$_{20}$ (2.38 $K$) and EuTi$_2$Al$_{20}$ (3.2 $K$) which is indicating a thermal arrest due to the latent heat.  The temperature variation of $S_{mag}$  for GdTi$_2$Al$_{20}$ saturates to a value $0.95R\ln8$  while the other magnetic systems exhibited still lower entropy saturation values in the high temperature limit. ($C_P-\gamma T)/T^3$ vs $T$ plot showed a maximum near 27 $K$ for all the compounds indicating the presence of low frequency Einstein modes of vibrations. Resistivity measurements showed that all the samples behave as normal Fermi liquid type compounds and $\rho(T)$ due to electron-phonon scattering follows Bloch-Gr$\ddot{\textrm u}$neisen-Mott relation in the paramagnetic region.
	
\end{abstract}


\noindent 
\maketitle
\section{Introduction}
The magnetism of R-TM (R- Rare earth element; TM- Transition metals) inter-metallic compounds is considered to be very intriguing and an interesting problem in condensed matter for several decades \cite{HMM1,HPCR1,HMM2}. The interaction between the 4f/5f local moments with the conduction electrons results in many different exotic phenomena such as complex magnetic ordering \cite{Jensen}, heavy fermion behavior\cite{RMP}, Kondo lattice systems \cite{Coqblin}, unconventional superconductivity and quantum criticality etc. \cite{Gegenwart,Steglich,bauer}.   Frank-Kasper cage compounds are an interesting class of materials which are investigated extensively in literature \cite{Thiede,KACZ,Kanges} owing to their strong correlation behaviour, unusual thermoelectric properties \cite{Canfield1}, rattling phenomena of guest ions \cite{Hiroi} superconductivity and multipolar ordering \cite{Takabatake}. These compounds crystallize in the CeCr$_2$Al$_{20}$ type cubic structure with space group $Fd\overline{3}m$. In this structure, the rare-earth and transition metals atoms occupy the high symmetry Wyckoff positions viz $8a$ and $16 d$ sites respectively whereas Al atoms contains three sites ($16c, 48f$ and $96g$). The important feature of this structure is that both rare-earth and transition metal atoms are placed in an over sized Frank-Kasper polyhedron formed by Al atoms. The rare-earth atom is surrounded by 16 Al atoms consisting of 12 Al1 and 4 Al3 atoms whereas the transition metal is centered at the CN12 polyhedron formed by 6 Al1 and 6 Al2 atoms. \cite{Kanges,Thiede}. Among the family of CeCr$_2$Al$_{20}$ type compounds Yb and Pr ion bearing systems owing to their strong correlation, display heavy fermion behaviour and unconventional superconductivity \cite{Canfield2,Takabatake}.  Jia \textit{et al.}  have observed nearly ferromagnetic Fermi liquid behavior and high ferromagnetic ordering temperature on  (Y,Gd)Fe$_2$Al$_{20}$ compounds\cite{Canfield3}. Another interesting system belong to the 1-2-20 family would be the LaV$_2$Al$_{20}$ compound. Onosaka \textit{et al} have observed large diamagnetic susceptibility and unusual orbital contribution from magnetic measurements and ascribed the negative susceptibility to the additional orbital contribution from V-$\textit{3d}$ electrons and specific evolution of the electronic band structure \cite{onosaka}. Recently, Hirose \textit{et al.} has explained the reasons for large negative susceptibility through the analysis of magnetization and de Haas-van Alphen oscillation measurements \cite{Hirose}. In the case of rare-earth inter-metallic compounds the total angular momentum associated with the rare-earth ion is expected to have 2J+1 degeneracy (J- total angular momentum) and due to the non spherical 4f charge distribution, the rare-earth ion bearing compounds are prone to have strong crystalline electric field effect (CEF) and are often found to exhibit large magneto-crystalline anisotropy. However in the case of Gd and limited Eu cases the CEF is expected to be absent due to their pure spin  state (S) with zero orbital momentum (L = 0). It is important to note that the europium and gadolinium based compounds belonging to BaNiSn$_3$ type tetragonal, Ni$_2$In hexagonal and ThCr$_2$Si$_2$ orthorhombic structures often exhibit double magnetic transition\cite{Neeraj}, amplitude modulated magnetic structure \cite{Lemoine} and valence fluctuation \cite{Klaus}. The nature of the magnetism and electronic properties of europium and gadolinium based compounds in this family are not investigated thoroughly in literature. Thiede et al. have  reported the synthesis conditions, structural parameters and exploratory magnetic properties for various RTM$_2$Al$_{20}$ compounds \cite{Thiede}. It is observed in GdFe$_2$Al$_{20}$ and GdCo$_2$Al$_{20}$ by Jia et al. that a subtle variation in the electronic structure showed conspicuous difference in the magnetic properties. The compound GdFe$_2$Al$_{20}$ exhibits ferromagnetic ordering with remarkably high ordering temperature ($T_C$ = 86 K) whereas the GdCo$_2$Al$_{20}$ shows commensurate equal moment-AFM ordering below 5.7 K \cite{Canfield3}. In this report we present analyses of low temperature physical properties in detail for RTM$_2$Al$_{20}$ [R = Eu, Gd and La ; TM = V and Ti] compounds.
\section{\label{sec:level1}Experimental details}
\subsection{\label{sec:level2}Crystal Growth}
Single-crystalline samples were prepared by high temperature solution growth method. The nominal composition according to the stoichiometry formula 1:2:20 is added to an alumina crucible. In addition, excess aluminum flux is added in accord with 1:7.5 (charge:flux) weight ratio. The crucible was then sealed inside a double necked quartz tube (ampule) as shown in Fig.1b. The quartz tubes for the crystal growth are designed in such a way that both crystal-growth and the decanting of the excess flux could be executed at one heating cycle. This procedure is found to be effective due to the fact that removing aluminum excess flux in the second heating cycle poses difficulty. After closing one end of the quartz ampule we evacuated the tube to the pressure of $10^{-6}$ torr before finally sealing the tube.  Samples were heated to $1100\,^{\circ}\mathrm{C}$ at the rate of $30.6\,^{\circ}\mathrm{C}/hr$ using a CARBOLITE Box furnace and maintaining this temperature for 24 hrs.  The melt is then slowly cooled down to $775\,^{\circ}\mathrm{C}$ at the rate of $1\,^{\circ}\mathrm{C}/hr$. The quartz tubes were quenched in atmosphere with the help of a Kokusan H103N low speed centrifuge. The excess Al flux is spun off and collected in the bulb of the quartz ampoule.
\subsection{\label{sec:level2}Crystal structure and elemental analysis}
The crystal structure analysis was carried out by X'Pert Pro x-ray diffractometer employing Cu-K$\alpha$ radiation. The observed x-ray pattern was refined by GSAS software to extract crystallographic information \cite{EXPGUI,GSAS}. The chemical composition and phase purity analysis are done by x-ray micro analysis using a Cameca SX100 Electron probe micro analyzer(EPMA). The electron micro probe analysis was performed by  measuring the intensities of the L$\alpha$ lines of the rare-earth elements, the K$\alpha$ lines of the transition metals and the second-order K$\alpha$ line of aluminum with wavelength-dispersive spectrometers. The probe was calibrated on rare-earth orthophosphates and pure metals, respectively \cite{jarosewich,donovan}. Quantification of the stoichiometry is done by measuring the net intensities of the respective characteristic X-ray lines of 20 randomly selected points. Laue back reflection diffraction is done using the Huber back-reflection Laue machine with x-ray wavelength $\delta\lambda$ = (0.5-4.2) \AA.  The Laue pattern is recorded on an image plate which was located along the camera axis with 3-4 cm distance away from the crystals. The recorded image was digitized with the help of an image plate scanner interfaced with a computer. After identifying the orientation, the crystals were cut by a electric discharge wire cutting machine as a well shaped parallelepiped suitable for physical property measurements with approximate dimensions 2-6$\times$  1-2$\times$1-1.5 (in mm). 
\subsection{\label{sec:level2}Physical Properties}
The magnetic measurements were carried out between 2 K to 300 K using the commercial MPMS-SQUID 7XL Quantum Design Magnetometer and Physical Property Measurement system with VSM attachment. Specific heat (C$_p$)  measurements were done by quantum design PPMS 9T machine using helium-4 heat capacity option. PPMS employs two-$\tau$ relaxation method to extract heat capacity from a sample under quasi-adiabatic conditions. The sample is stuck and thermally coupled using Apiezon N grease whose heat capacity is initially measured in the same temperature range in which sample C$_p$ is to be measured. The heat capacity measurements were extended down to 360 mK using the $^3He$ insert option from Quantum Design, Inc. The resistivity measurements were done using conventional linear four probe method with the dc transport option of PPMS system.
\section{Results and Discussion}
\subsection{Structural and elemental analysis}
Fig.1a depicts the essential features associated with the CeCr$_2$Al$_{20}$ type structure.  The detailed discussion about the structure can be found elsewhere in literature \cite{Thiede,KACZ}. Here we emphasize that all the compounds crystallize in cubic structure with space group $Fd\overline{3}m$ (Refer Fig.2). We indexed the x-ray pattern with allowed reflections belonging to the CeCr$_2$Al$_{20}$ type structure with an tiny additional peak positioned at the 2 $\theta$ value  $ \sim 39.1\,^{\circ}$ (See inset of Fig.2). The additional peak is corresponding to the FCC Al phase. It is to be noted here that the microprobe studies showed only the primary 1-2-20 phase when observing back scattered electron image (picture not shown) and the analysis indicates that the interior portion of the single crystals which was selected for physical property measurements, are free from elemental Al phase. The refined structural parameters, R-R distance and the occupancy values are given in the Table 1. The refined lattice parameter values for LaTi$_2$Al$_{20}$  is observed to be 14.7621(1) \AA\ and for  LaV$_2$Al$_{20}$ the value is 14.6130 (2) \AA. These values agree with the previously reported lattice parameter for these compounds \cite{onosaka}. It is interesting to note the lattice parameter anomaly between the europium and gadolinium compounds (Refer Fig.3). Figure 3 shows the variation of lattice parameter with respect to the rare-earth elements for the compounds RTM$_2$Al$_{20}$ (R = Eu, Gd and La; TM = V and Ti) along with selected iso-structural compounds reported in Ref. \cite{Kanges,Thiede,niemann}. The deviation from lanthanide contraction is due to the divalent electronic state of europium ion as opposed to the trivalent electronic state for other RTM$_2$Al$_{20}$ (R = Gd, La; TM = V and Ti) compounds. Even though the lattice parameter anomaly is observed by Thiede et al \cite{Thiede} in the Europium compounds, for the specific system EuV$_2$Al$_{20}$ we observed large deviation in the lattice parameter values. Since Thiede et al has followed different growth conditions the deviation in the lattice parameter could be due to the a small variation in the stoichiometry \cite{Thiede}. Another interesting aspect of our structural analysis is the lattice contraction between vanadium and titanium derivatives. The non-magnetic compounds LaTi/V$_2$Al$_{20}$ showed 3\% volume change between vanadium and titanium compounds which is in concert with the observation by Kanges et. al \cite{Kanges}. However notable difference is observed with the magnetic compounds in which the volume contraction is only 2.5\% for Gd compounds whereas europium compounds showed 0.5\% volume change between the V and Ti compounds.
\begin{figure}[!h]
	\centering
	\includegraphics[scale=0.33]{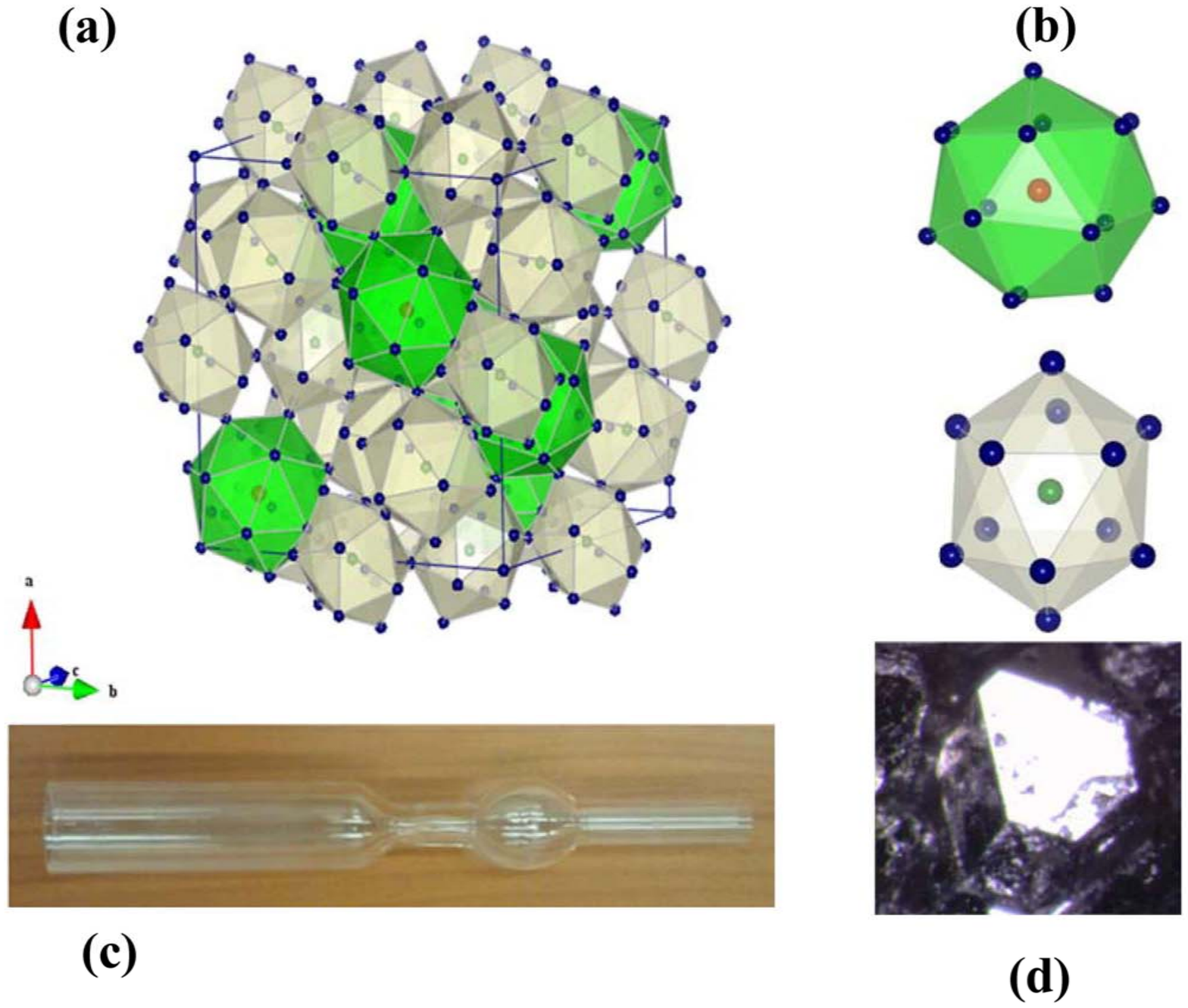}
	\caption{(Color online)(a) The unitcell of RTM$_2$Al$_{20}$ type compounds, showing interpenetration of connected polyhedra of R and TM atoms. The R, TM and Al atoms are shown as red, green and  blue spheres respectively. (b) The coordination polyhedra for the R and TM atoms. The rare-earth sublattice is formed by corner sharing 16 coordinate polyhedra whereas the transition metal atom is surrounded by 12 Al atoms which constitute a corner sharing distorted icosahedron. (c) the double necked quartz tubes employed for the crystal growth process. (d) representative single crystal of RTM$_2$Al$_{20}$  clearly showing the $<111>$ orientation and principal growth facet.} 
\end{figure}

\begin{figure}[h!]
	\centering
	\includegraphics[scale=0.36]{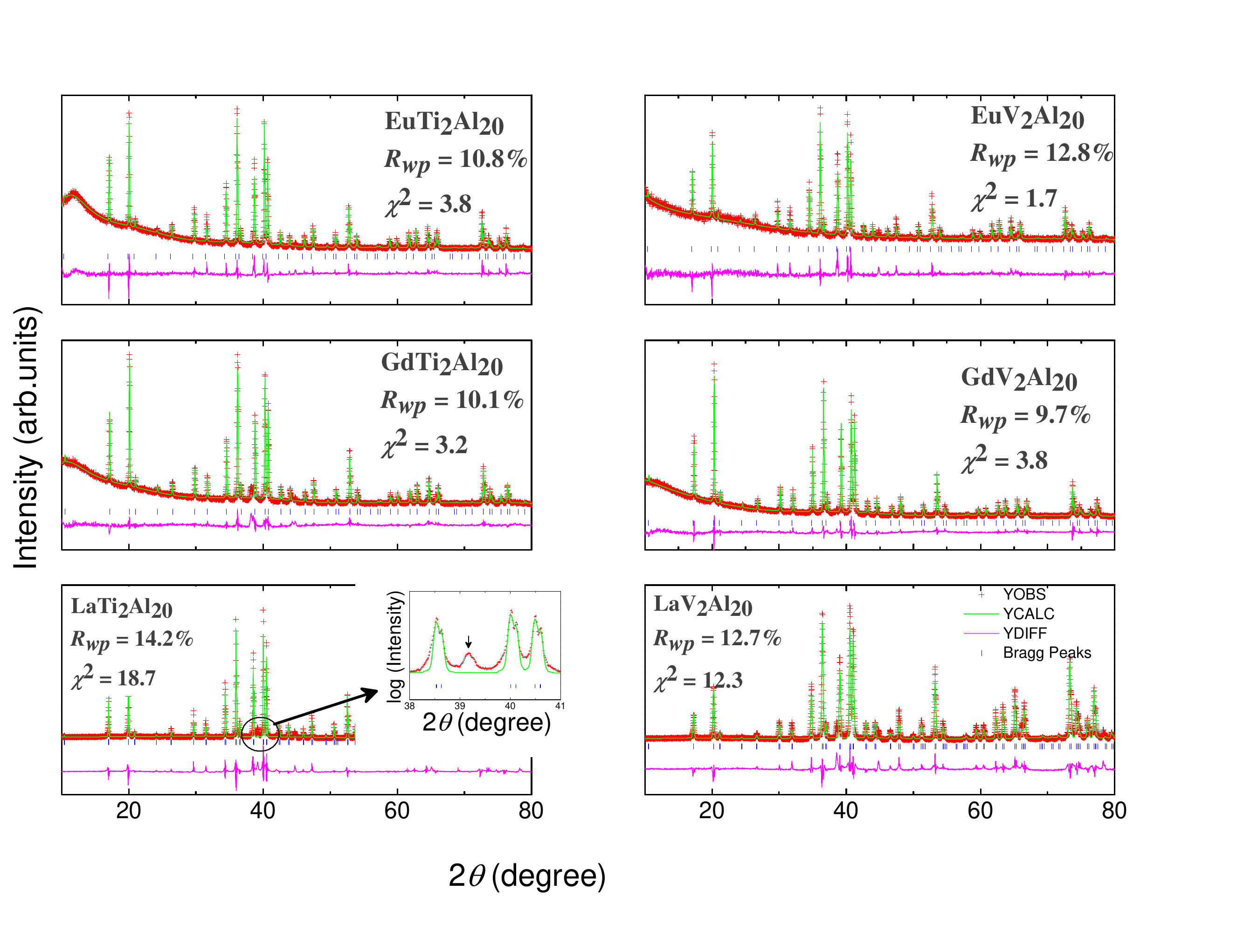}
	\caption{(colour online) Rietveld refinement of powder X-ray diffraction patterns modeled using $Fd\overline{3}m$ space group. The red plus symbols represent the experimental data whereas calculated pattern is showed as solid green line. The difference pattern between the fit and experimental data is given as pink solid line. Blue vertical bars represent the allowed Bragg peak positions for the CeCr$_2$Al$_{20}$ type structure. The inset in the bottom-left panel shows enlarged x-ray pattern for the compound LaTi$_2$Al$_{20}$ and the unaccounted peak represents the residual Al phase (JCPDS: 00-001-1176)
		\newline $R_{wp}=[\frac{\sum w_i(y_{io}-y_{ic})^2}{\sum w_i(y_{io})^2}]^{1/2}$; $\chi^2 = [\frac{R_{wp}}{R_{exp}}]^2$;
		\newline $y_{io}$, $y_{ic}$ represent observed and computed intensity values respectively. $w_i$ indicate the weight factor.}.
	
\end{figure}

\begin{figure}[htb!]
	\centering
	\includegraphics[scale=0.3]{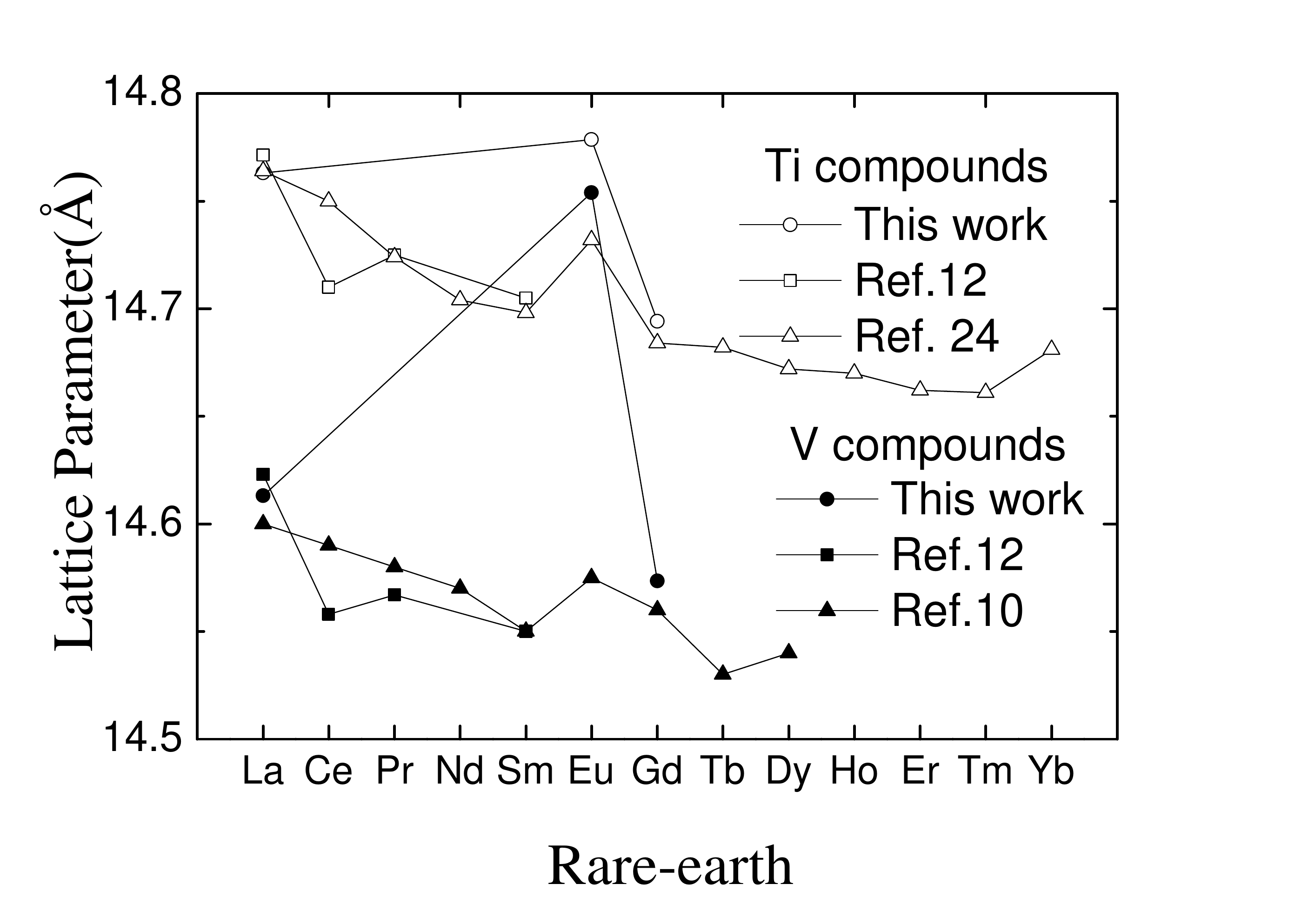}
	\caption{Lattice parameter variation as function of rare-earth elements for the compounds  RTM$_2$Al$_{20}$ (M = Ti and V). Solid and open circles are corresponding to the Ti and V derivatives respectively. } 
\end{figure}

The inter-atomic distances deduced from the refined structural parameters are  typical for this structure system and comparable to the other iso-structural compounds viz. GdCo$_2$Zn$_{20}$ and GdFe$_2$Zn$_{20}$ \cite{Canfield2}. The nearest and next nearest neighbour for the rare-earth is surrounded by only Al atoms. The shortest inter-atomic distance (d$_{R-Al1}$) for the rare-earth varies between 3.15 to 3.25  $\textrm\AA$ for different rare-earth elements.  Finally, the occupancy of the rare-earth site was observed to be somewhat lower than 100 \% indicating possible rare-earth deficiency (Table 1). Even though our refinement showed presence of rare-earth deficiency, quantification of it using conventional powder x-ray diffraction measurements and Rietveld refinement may not be accurate. Hence, the quantification of the individual elements and estimation of the stoichiometry are done by micro probe analysis. Atomic percent of individual elements, estimated stoichiometry and recalculated molecular weight values are given in Table 2. The compounds (La/Eu)V$_2$Al$_{20}$ showed $\sim$10 \% of rare-earth deficiency and $\sim$ 8 \% site vacancy is estimated for the compound GdTi$_2$Al$_{20}$. Other compounds showed nominal composition within the accuracy of the EPMA measurements (about $\pm 2\%$). 
\begin{table*}[!t]
	\caption{Atomic coordinates,  occupancy and lattice parameter values of cubic  {RTM$_2$Al$_{20}$ R= Gd, Eu; TM = V and Ti} compounds  extracted from Rietveld refinement. Standard deviations of the refined parameters are given in the parentheses}
	\label{table:1}
	\setlength{\tabcolsep}{6pt}
	\begin{tabular}{c c c c c c c c c }
		\hline
		\hline
		&  &  &  &  & & & & \\
		& Compound & Atoms &  $x/a$ & $y/a$ & $z/a$ & Occupancy & Bond lengths (\AA)  &\\
		& & & & &  & & R-R &   \\
		& & & & & & &  &     \\
		\hline 
		& & & & & & &  &     \\			
		&	& Eu  & 1/8 & 1/8 & 1/8 &  0.98(6) &  &    \\
		&	& Ti  & 0.5 & 0.5 & 0.5 &  0.97(1) &  &    \\
		& EuTi$_2$Al$_{20}$	& Al1  & 0 & 0 & 0 &  1.01(2) & 6.39914(10) &     \\
		& a = 14.7785 (3)\AA & Al2  & 0.486(4) & 1/8 & 1/8 &  0.99(2) &  &    \\
		&	& Al3  & 0.059(5) & 0.059(3) & 0.325(3) &  1.01(1) &  &    \\
		
		& & & & & & &  &   \\
		
		&	& Eu  & 1/8 & 1/8 & 1/8 &  0.86(7) &  &    \\
		&	& V  & 0.5 & 0.5 & 0.5 & 0.88(15) &  &    \\
		& EuV$_2$Al$_{20}$	& Al1  & 0 & 0 & 0 &  1.09(14) &  6.3884(10) &  \\
		& a = 14.7539 (3)\AA	& Al2  & 0.486(5) & 1/8 & 1/8 & 1.05(20) &  &    \\
		&	& Al3  & 0.060(5) & 0.060(3) & 0.325(3) & 1.10(1) &  &    \\
		& & & & & & &  &    \\
		&	& Gd  & 1/8 & 1/8 & 1/8 &  0.98(6) &  &   \\
		&	& Ti  & 0.5 & 0.5 & 0.5 &  0.97(1) &  &    \\
		& GdTi$_2$Al$_{20}$	& Al1  & 0 & 0 & 0 &  1.01(2) &  6.3604(8) & \\
		& a = 14.6941 (2)\AA	& Al2  & 0.485 (3) & 1/8 & 1/8 &  0.99(2) &  &    \\
		&	& Al3  & 0.059 (3) & 0.059(2) & 0.325(6) &  1.01(1) &  &   \\
		
		& & & & & & &  &    \\
		&	& Gd  & 1/8 & 1/8 & 1/8 &  0.96(6) &  &    \\
		&	& V  & 0.5 & 0.5 & 0.5 &  1.09(12) &  &     \\
		& GdV$_2$Al$_{20}$	& Al1  & 0 & 0 & 0 & 1.00(2) &   6.3152 (9) &  \\
		& a = 14.5735 (2)\AA	& Al2  & 0.485 & 1/8 & 1/8 & 1.17(2) &  &    \\
		&	& Al3  & 0.059(2) & 0.059(2) & 0.326(2) & 1.01(1) &  &   \\
		
		& & & & & & &  &   \\
		&	& La  & 1/8 & 1/8 & 1/8 & 0.89 (5)   &  &    \\
		&	& Ti  & 0.5 & 0.5 & 0.5 &   1.02 (12)  &   &   \\
		& LaTi$_2$Al$_{20}$	& Al1  & 0 & 0 & 0 & 1.21 (25) & 6.3922 (6) &    \\
		& a = 14.7621 (1)\AA	& Al2  & 0.485 (3) & 1/8 & 1/8 & 1.07 (15) &  &    \\
		&	& Al3  & 0.058 (2) & 0.058 (2) & 0.326 (3) &  1.10 (10) &  &   \\
		
		& & & & & & &  &   \\
		&	& La & 1/8 & 1/8 & 1/8 &  0.76 (5) &  &   \\
		&	& V  & 0.5 & 0.5 & 0.5 &   0.89 (10) &  &    \\
		& LaV$_2$Al$_{20}$	& Al1  & 0 & 0 & 0 &  1.07 (3) & 6.3274 (10)  &     \\
		& a = 14.6130 (2)\AA	& Al2  & 0.485 (2) & 1/8 & 1/8 &  0.97 (12) &  &    \\
		&	& Al3  & 0.059 (2) & 0.059 (2) & 0.326 (3) &  1.025 (8) &  &   \\
		\hline
	\end{tabular}    
\end{table*}

\begin{table*}
	
	\caption{Chemical composition estimated from the net intensities of characteristics X-ray lines (EPMA-WDS). The recalculated molecular weight is represented as MW$^R$.}
	
	\label{table:1}
	\setlength{\tabcolsep}{6pt}
	
	\begin{tabular}{c c c c c c c c}
		\hline
		\hline
		&  &  &  &  & & & \\
		& Compound &  &  Atomic Percentage &  & Stoichiometry & MW$^R$ & $\frac{\textnormal {At.\% (Al)}}{\textnormal{At.\% (R)}}$ \\
		&  &  &  &  & & & \\
		& & R &  TM & Al & & & \\
		
		& & & & & & & \\
		\hline 
		& & & & & & & \\			
		& GdTi$_2$Al$_{20}$	& 4.03  & 9.21 & 86.74 & Gd$_{0.92}$Ti$_{2.11}$Al$_{19.95}$ & 784.42 & 21.3  \\
		& GdV$_2$Al$_{20}$	& 4.32  & 9.69 & 85.98 & Gd$_{0.99}$V$_{2.22}$Al$_{19.77}$ & 807.04 & 19.7 \\
		& EuTi$_2$Al$_{20}$	& 4.30  & 9.45 & 86.24 & Eu$_{0.98}$Ti$_{2.12}$Al$_{19.80}$  & 791.15 & 20.2 \\
		& EuV$_2$Al$_{20}$	& 3.93  & 9.43 & 86.62 & Eu$_{0.90}$V$_{2.12}$Al$_{19.92}$ & 782.23 & 22.1 \\
		& LaTi$_2$Al$_{20}$	& 4.26  & 9.42 & 86.32 & La$_{0.97}$Ti$_{2.16}$Al$_{19.85}$ & 773.71 & 20.2 \\
		& LaV$_2$Al$_{20}$	& 3.98  & 9.17 & 86.84 & La$_{0.91}$V$_{2.12}$Al$_{19.97}$ & 770.58 & 21.9 \\
		
		& & & & & & & \\

		\hline
	\end{tabular}
\end{table*}
\subsection{Magnetic susceptibility} 

Firstly, we will discuss the magnetic susceptibility of the two non-magnetic compounds La(V/Ti)$_2$Al$_{20}$. The two systems showed different temperature dependent susceptibility behavior (see Fig.4). The compound LaTi$_2$Al$_{20}$  exhibits temperature independent Pauli paramagnetism  in the temperature of investigation. In contrast, temperature independent diamagnetic behavior is observed for the compound LaV$_2$Al$_{20}$  in the high temperatures and exhibits an upturn towards positive regime below 10 K (Fig.4). 
\begin{figure}[h!]
	\centering
	\includegraphics[scale=0.36]{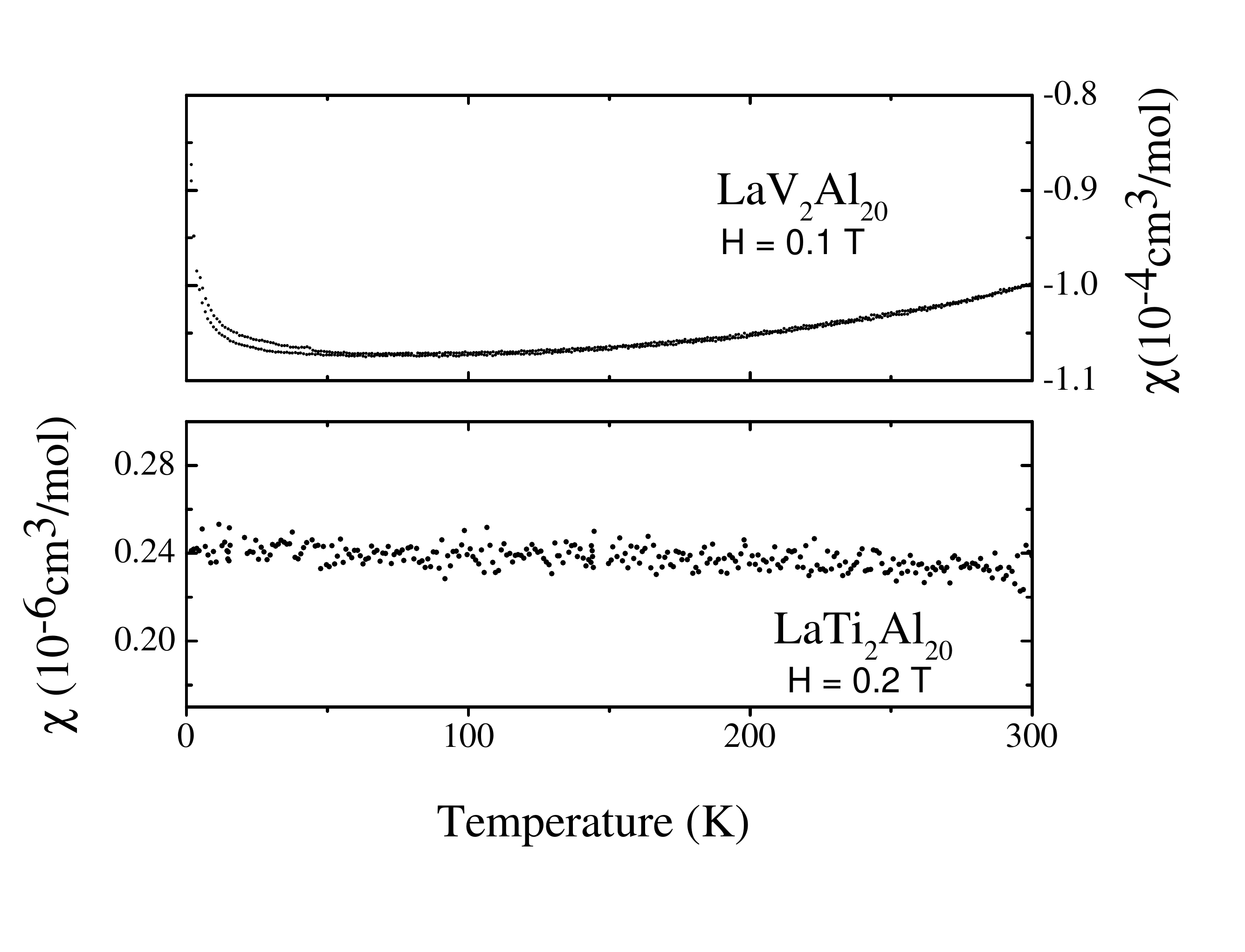}
	\caption{Temperature variation of susceptibility for the non-magnetic compounds La(Ti/V)$_2$Al$_{20}$. The volume
		susceptibility for the compounds calculated using
		the formula $\chi_V = (\rho/W)M$; where $W$ and $\rho$ are mass and density of the sample respectively.} \label{Figure5}
\end{figure}			

The magnetic susceptibility of any non-magnetic metallic sample is expected to have two contributions viz. orbital contribution arising from core electrons and paramagnetic contribution due to conduction electrons which is known as Pauli spin susceptibility \cite{onosaka}. Depending upon the strength of these contributions the temperature variation of any non magnetic susceptibility would follow different temperature variation. The total susceptibility  for a non magnetic metal can be written as the following expression
\newline
$\chi= \chi^{Pauli}+\chi^{orb} \qquad \qquad \textrm{(1)}$ 
\newline 
Where $\chi^{Pauli}$ is  the susceptibility due the conduction electrons and the value directly proportional to the total density of states at the Fermi Level.
\newline
$\chi^{Pauli}$ = $g^2/4\mu_B^2N(E_F). \qquad \qquad \textrm{(2)}$
\newline
where $g$ and $N(E_F)$ are spectroscopic splitting factor and density of states (DOS) at the Fermi level respectively.
The relationship between the Sommerfeld heat capacity coefficient and density of states can be written as the following expression. 
\newline
$\gamma=2/3\pi^2k_B^2N(E_F). \qquad \qquad \textrm{(3)}$
\newline
$N(E_F)$ expressed in states/eV.f.u and the electronic specific heat coefficient ($\gamma$) is in the units of eV/mol.K$^2$. 
\newline By comparing the two equations mentioned above the Pauli spin susceptibility can be written as 
\newline 
$\chi^{Pauli}$ = (1.357$\times$10$^{-5})N(E_F). \qquad \qquad \textrm{(4)}$ 
\newline
The Pauli spin susceptibility for LaTi$_2$Al$_{20}$ and LaV$_2$Al$_{20}$ was calculated to be 2.84$\times$10$^{-4}$ cm$^3$/mol and 2.93$\times$10$^{-4}$ cm$^3$/mol respectively. 
The second term in equation (1) comprises 3 contributions viz., Larmor  diamagnetic contribution, Landau-Peierls orbital term and  Van Vleck paramagnetic contribution. The Larmor's diamagnetic susceptibility was estimated by considering 3+ valence state for La ions and most stable valance states for Ti (4+) and V (5+) ions \cite{Susceptiblity}. The values are -70$\times$10$^{-6}$ cm$^3$/mol and -74$\times$10$^{-6}$ cm$^3$/mol for LaTi$_2$Al$_{20}$ and LaV$_2$Al$_{20}$ respectively. Even though the Pauli contribution is larger compared to the core diamagnetic contribution we see over-compensation of negative susceptibility in these two compounds La(Ti/V)$_2$Al$_{20}$. The additional contribution is expected to arise from the Landau-Peierls orbital contribution which can be related to the Pauli spin susceptibility by the following relation 
\newline
$\chi^{LP}$ = -($\frac{m_e}{3m^*}$) $\chi^{Pauli} \qquad \qquad \textrm{(5)}$
\newline
where $m_e$ and $m^*$ are free electron mass and electron effective mass respectively.
Onosaka et al. have given plausible reasons for the large negative susceptibility in the compounds LaV$_2$Al$_{20}$ and YV$_2$Al$_{20}$ by varying the number of valence electrons per formula unit through substituting Al and Sc in the 8a site. They argued that the sign and magnitude of total susceptibility for the compounds LaV$_2$Al$_{20}$ (-7.44$\times$10$^{-4}$ cm$^3$/mol), YV$_2$Al$_{20}$ (-1.9$\times$10$^{-4}$ cm$^3$/mol) and (+3.14$\times$10$^{-4}$ cm$^3$/mol) at 10 K showed large differences albeit having same electron count and comparable $\gamma$ values. Their finding suggest that additional orbital contribution from V-$\textit{3d}$ states and specific evolution of the band structure might be the reasons for high negative susceptibility in LaV$_2$Al$_{20}$ and YV$_2$Al$_{20}$ compounds \cite{onosaka}. Recently, Hirose et al. have carried out magnetization and de Haas Van Alphen studies on the single crystalline compounds LaV$_{(2-x)}$Ti$_x$Al$_{20}$ (x = 0; 0.05, 0.1, 0.2) and explained that the compound LaV$_2$Al$_{20}$ shows a small hole like Fermi surface and possesses extremely small effective mass 0.067 $m_e$ which  are responsible for large diamagnetic susceptibility \cite{Hirose}. Our susceptibility measurements corroborates the analysis by Hirose et al. that additional contribution from V-$\textit{3d}$ states could lead to enhancement of Landau-Peierls contribution.
\\
In order to observe the magnetic state and electronic configuration of the Eu/Gd ions, susceptibility and magnetization measurements were done for the magnetic samples by applying field along $<111>$ direction. The temperature variation of the dc magnetization and inverse susceptibility is presented in Fig.5 and the data show that all 4 of the Eu and Gd compounds order magnetically in a spin arrangement that is plausibly anti-ferromagnetic. But in addition, the susceptibility undergoes a discontinuous jump below $T_N$ for the Ti compounds at $T_o$ = 2.38 K and $T_o$ = 3.23 K for GdTi$_2$Al$_{20}$ and EuTi$_2$Al$_{20}$ respectively, which is an indication of possible double magnetic transition and complex spin arrangement in these systems. The sudden jump in the magnetization for the Ti compounds may be attributed to the first-order type magnetic transition. Within the accuracy of our magnetic measurements the susceptibility did not show hysteresis behaviour with cooling and warming protocol (see figure 5a and 5c). However, clear evidence of first-order nature of the phase transition is seen in the heat capacity measurements and the data is discussed in  the section 3.4. Further, the compound EuV$_2$Al$_{20}$ showed an upturn near 2.8 K and tendency of saturation when the temperature is reduced  which signals a possible ferromagnetic transition below the N{\'e}el temperature. For the specific case of EuTi$_2$Al$_{20}$  susceptibility and magnetization measurements were carried out by applying field along 3 different crystallographic directions ($<100>, <110> \textrm {and} <111>$) and the measurement showed no appreciable anisotropy in the magnetic susceptibility and magnetization measurements (See Fig.6).  Since the magnetic susceptibility measurements along the principal axes  could not resolve the magnetic easy axis and anti-ferromagnetic ordered moment direction we refrain from drawing conclusive remarks on the magnetic structure of these compounds. 
\begin{figure}[h!]
	\centering
	\includegraphics[scale=0.36]{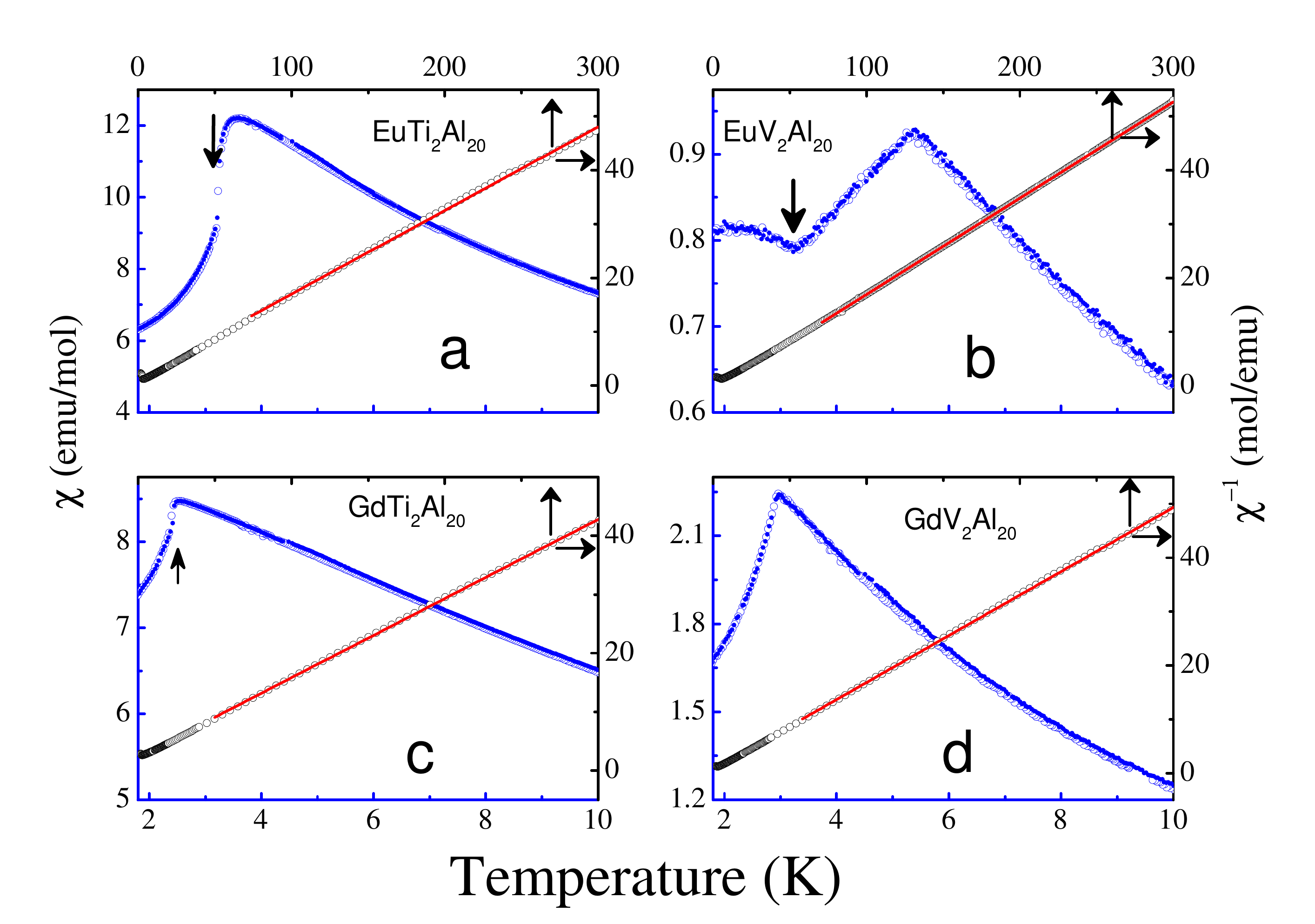}
	\caption{(Color online) panels (a-d) represent temperature variation of both $\chi$ and $\chi^{-1}$. Solid and open blue symbols represent the susceptibility data (H = 20 Oe along $<111>$)  carried out in cooling and warming protocol. The arrows show the possible double magnetic transition below $T_N$. Black open symbols represents the inverse susceptibility data (H = 1000 Oe $<111>$) where the red lines indicate the CW fitting as explained in the text.}
\end{figure}

\begin{figure}[htb!]
	\centering
	\includegraphics[scale=0.33]{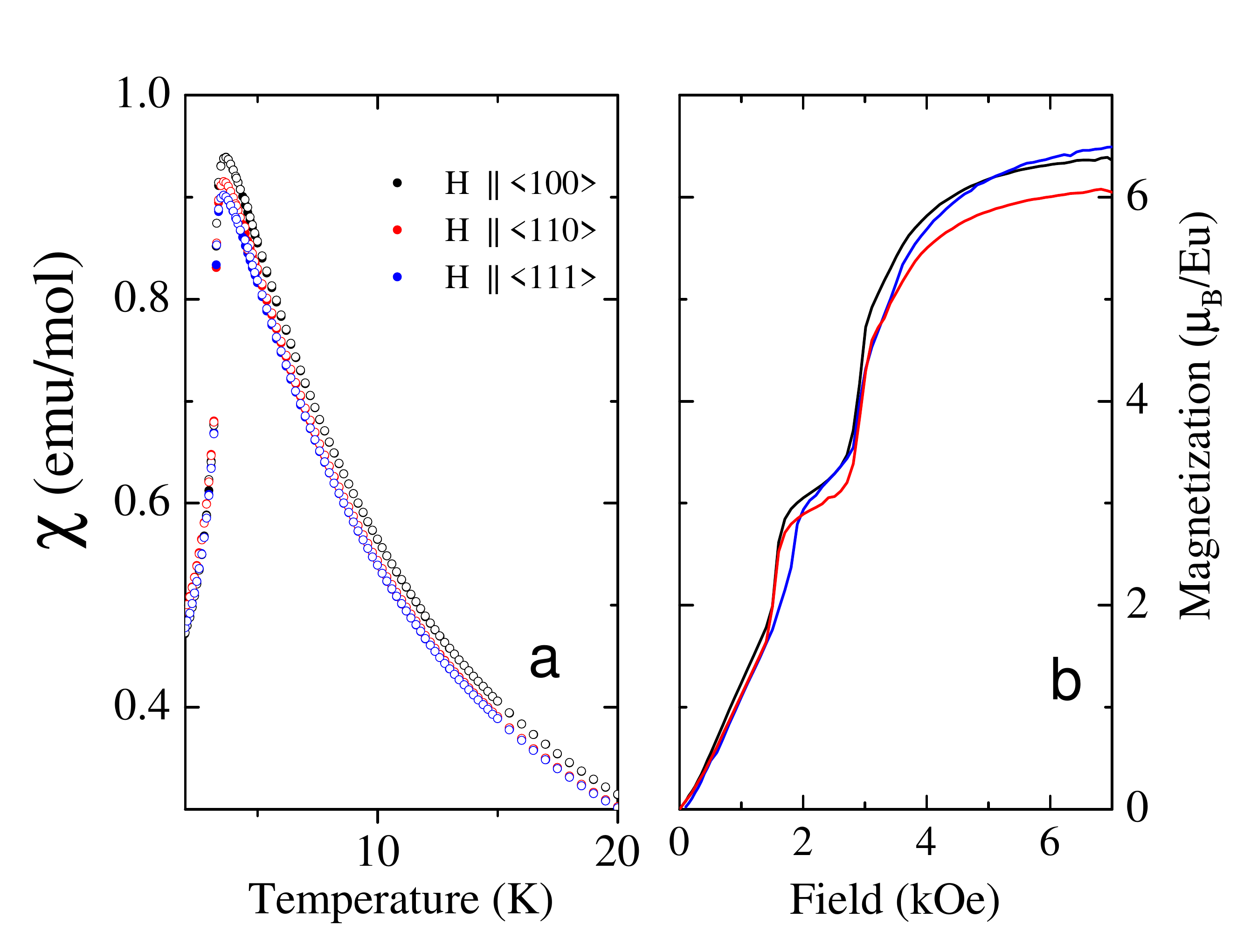}
	\caption{(Color online) left : Temperature variation of magnetic susceptibility  along three principal crystallographic direction $<100>$, $<110>$ and $<111>$ for the compound EuTi$_2$Al$_{20}$ (b) Magnetization data recorded at 2 K when the field is applied along the above mentioned crystallographic directions to show anisotropic behaviour}
\end{figure}
First-order transition is occasionally observed in R-TM intermetallic compounds and transition metal alloys, for example the compounds Gd$_5$(Si$_{1-x}$Ge$_x$)$_4$ x $<$ 0.5 are  well known room temperature magnetocaloric material which undergo a ferromagnetic-paramagnetic transition coupled to orthorhombic-monoclinic crystallographic phase transformation. Apart from the above mentioned compound the first-order magnetic transition is observed in the following compounds; Sm$_2$IrAl$_8$(AFM to Para $T_N$ = 14.2 K), MnAs (FM-Para $T_c$ = 307 K) and RCo$_2$ (R = Er, Dy and Ho; Ferrimagnetic to Paramagnetic) \cite{SmIr,BR,Ranke}. Within the family of CeCr$_2$Al$_{20}$ type compounds the first-order magnetic transition is observed for the first time, and thus far only in the compounds (Eu/Gd)Ti$_2$Al$_{20}$. The important parameters to study the phase separation between the first-order and second-order transitions are deformation dependent exchange energy (T$_N$) and the magnetic ion inter-atomic spacing spacing (R-R distance). In the case of our magnetic samples we observed  0.2 percent of R-R spacing change between two Eu compounds and $\sim$ 0.8 percent change for the Gd compounds (Table 1). For the corresponding change in the interatomic spacing we observed small change ($\sim 2-3 $ K) in the transition temperature between V and Ti derivatives (Refer Table 3). Further measurements viz thermal expansion, compositional dependence of T$_N$ and high pressure measurements are needed to study microscopic origin of first order transition in the (Eu/Gd)Ti$_2$Al$_{20}$ as discussed in the case of Gd$_5$(Si$_{1-x}$Ge$_x$)$_4$ and  Mn$_{1.1}$Fe$_{0.9}$P$_{1-x}$As$_x$ \cite{BR_Gd5Si2Ge2,MnP}. 
\\
\begin{center}
\begin{table}[!t]
	\caption{\label{tab1} Summary of magnetic properties of RTM$_2$Al$_{20}$ R= Gd, Eu; TM = V and Ti compounds. The parameters listed are: N{\'e}el temperature $T_N$, Curie constant  C, Weiss temperature $\theta_P$, effective magnetic moment $\mu_{eff}$, Saturation magnetization $M_{Sat}$}, Number of magnetic ions (estimated) $N$
	\setlength{\tabcolsep}{6pt}
	\centering
	\begin{tabular}{c c c c c c c c}
		\hline
		\hline
		& & & & & & & \\ 
		
		& Samples &  $T_N$ & $C$ & $\theta_P$&$\mu_{eff}$ & $M_{sat}$  & $N$ \\
		& & (K) & $\frac{\textrm {emu.K}}{\textrm {mol.Oe}}$  &(K) & $\mu_B$ & $\mu_B$ &  \\ 
		
		\hline
		& & & & & & &   \\ 
		& EuTi$_2$Al$_{20}$  & 3.6 & 6.4 & -9.2 & 7.15 & 5.53 & 82.3  \\
		& & & & & & &   \\
		& EuV$_2$Al$_{20}$  & 5.5 & 5.54 & -5.5 & 6.65 & 5.10 & 70.3 \\
		& & & & & &  &  \\
		& GdTi$_2$Al$_{20}$ &  2.6 & 7.4 & -17.9 & 7.69 & 6.16 & 94  \\
		& & & & & &  &  \\
		& GdV$_2$Al$_{20}$ & 3.1 & 6.15 & -3.54 & 7.04 & 5.2 & 78 \\
		& & & & & &  &   \\
		
		\hline
	\end{tabular}
\end{table}
\end{center}
The temperature variation of the susceptibility is fitted with following relation $\chi=C/(T-\theta_p)$ in the temperature range between 50 K to 300 K. In the above expression  C and $\theta_{p}$ stands for Curie-Weiss (CW) constant and paramagnetic Curie temperature respectively. Further the relation between Curie constant and effective magnetic moment $\mu_{eff}$ can be written as $C = N\mu_{eff}$/3$k_{B}$ where $N$ = total number of magnetic ions per formula unit.  From the CW fitting  Eu effective moment was observed to be 7.15 (2) $\mu_{B}$  for the case of EuTi$_2$Al$_{20}$ and for the vanadium counter part the value is 6.36 (7) $\mu_{B}$ (Table 3). The effective magnetic moment values are smaller than the theoretical value for divalent europium ion (7.94 $\mu_{B}$ $^8S_{7/2}$). In the case of Gd samples the $\mu_{eff}$ amounts to 7.69 (2) $\mu_{B}$  for Ti sample whereas the GdV$_2$Al$_{20}$ compound showed lower value of 7.04 (1) $\mu_{B}$ which may be compared with the Gd free-ion value of 7.94 $\mu_{B}$. The structural and elemental analysis indicated a possible small rate of rare-earth vacancy  in the compounds however the estimated rare-earth deficiency does not scale with the full moment deficit  observed from CW analysis. By assuming Eu(2+) valence state the estimated rare-earth vacancy from the CW fitting is 30 $\%$ for the compound  EuV$_2$Al$_{20}$ whereas the compositional analysis suggest a rare-earth deficiency as high as 10 $\%$. Similar difference is observed in the case of  GdV$_2$Al$_{20}$ as well. By considering Eu(2+) and Gd(3+) valence states and cubic point symmetry, the reduced effective moment is not clear yet in the present studies. However it should be noted that in high temperature flux-growth method incorporation of a small amount of flux materials in the interstitial sites and grain boundaries cannot be avoided. Another interesting observation from the magnetic susceptibility measurements is the larger deviation of free ion moment between Ti and V derivatives (See Table 3). Even though LaV$_2$Al$_{20}$ compound showed a temperature independent diamagnetic behaviour, in low temperatures the susceptibility indicated a possible CW behaviour (See Fig.3)of free V-ions hence in the case of V based compounds GdV$_2$Al$_{20}$ and EuV$_2$Al$_{20}$ additional contribution from V moments and its CW behaviour should also be taken into account. We attempted to extract the V moment after carefully subtracting the diamagnetic contribution but temperature range in which the CW behaviour was observed only 2-10 K range hence the fitting resulted in unphysical values.
\subsection{Magnetization}

The single crystal magnetization data in the temperature range between 2 K to 6 K is presented in figure 6. Isothermal magnetization at 1.8 K for the compound EuTi$_2$Al$_{20}$  increases linearly till H = 1.5 T, typical for simple AFM spin arrangements. The magnetization showed a sudden jump at 1.5 T, followed by two successive metamagnetic jumps at the field value of 2.8 T and 2.9 T (Refer panel (a) of Fig.7). 
\begin{figure}[h!]
	\centering
	\includegraphics[scale=0.35]{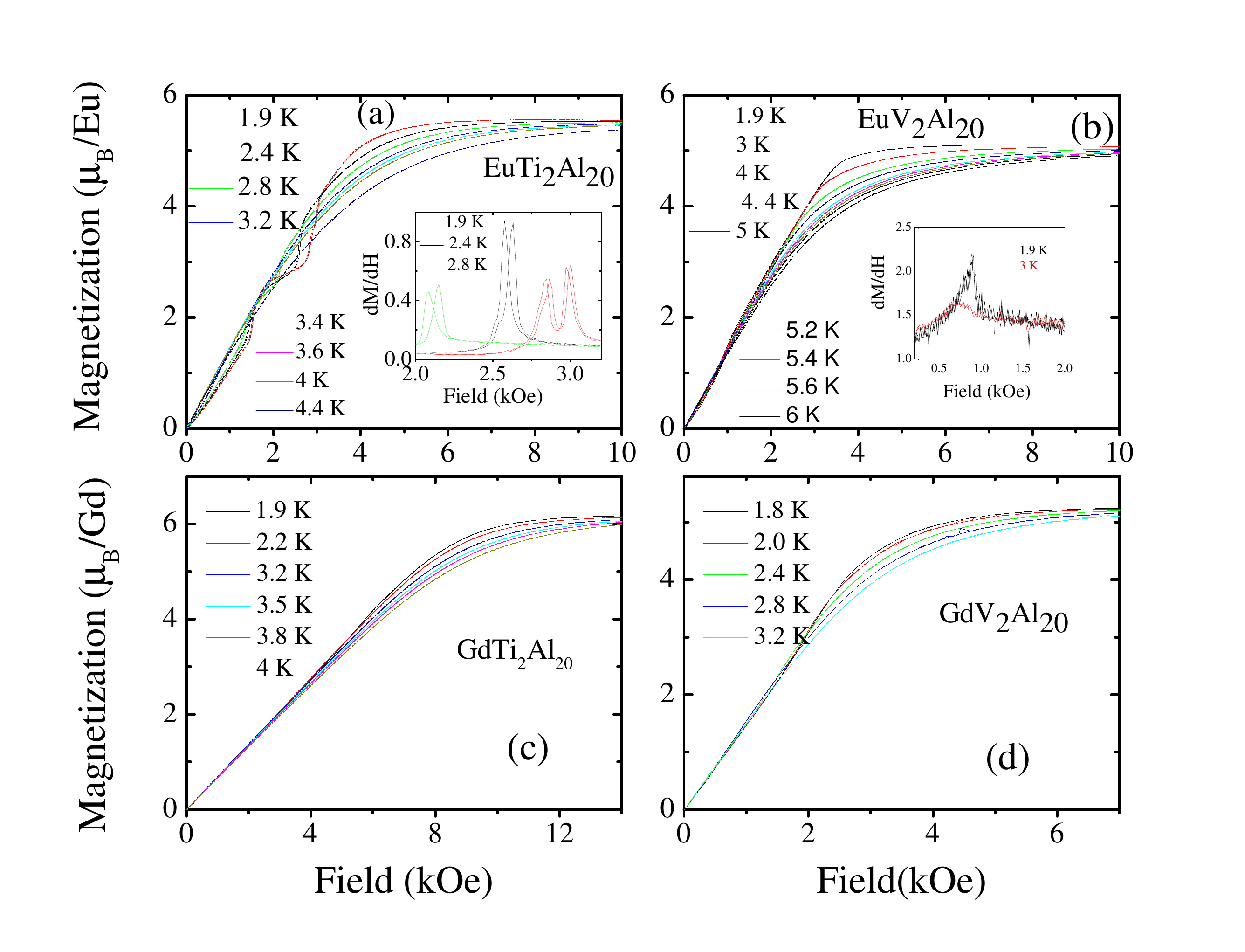}
	\caption{(Color online) (a-d) Isothermal magnetization measurement for the compounds RTM$_2$Al$_{20}$ (R = Eu, Gd; TM = V and Ti) in the temperature range between 1.8 K to 5 K. The field is applied along the $<111>$ direction. The inset of panel (a) and (b) shows temperature evolution of the metamagnetic transition plotted using the first-order derivative of the magnetization curves.}
\end{figure}
Then continuous rotation of the magnetic moment occurs before saturating in the fields above 6 T to the value of 5.53 $\mu_B$. The saturation magnetization is less than the expected value for Eu(2+) ion assuming g =2 and  S = 7/2 ($\mu_{sat}$ =  $7\mu_B$/Eu). By considering the Eu moments aligned in the $<111>$ direction and a weak anisotropic energy, the spin-flop transition occurs if the two sublattice moments flop to a perpendicular direction with respect to the easy magnetization axis or applied field direction. The evolution of spin-flop transition is shown in the inset of figure 6(a) through  first order derivative of magnetization with respect to field ($\frac{dM}{dH}$). As temperature is increased from 1.8 K to 3.6 K, the two metamagnetic transitions shift towards lower fields with considerable increasing in the broadening of the transition. From the derivative of the magnetization data it is clear that these transitions are first-order in nature due to the hysteresis behavior between increasing and decreasing field (Inset of fig6a). The isothermal magnetization below and above the transition temperature showed slightly different field variation for the compound EuV$_2$Al$_{20}$. Even though the magnetization showed a spin-flop transition at the field value of 1 T, the transition is not as prominent as in the case of EuTi$_2$Al$_{20}$ compound. The inset of Fig.7b shows the first order derivative of magnetization plotted against the applied field. The field variation $dM/dH$ at 1.8 K shows a weak linear function till the field value of 0.8 T and upon further increase in the field the derivative showed smeared peak centered around 1 T indicating a weak metamagnetic behaviour. These features are observed in other europium compounds as well, for example, Anand et al. has observed a smeared peak around 0.5 T in the $dM/dH$ vs $H$ plot for the orthorhombic EuCu$_{1.82}$Sb$_{2}$ compound \cite{VKAnand}. The reason could be due to the distribution of the critical field over a finite range opposed to single critical field or the transition could be a field induced second order transition. On the other hand, the isothermal magnetization of the Gd compounds follows a typical anti-ferromagnetic behaviour with no sign of any metamagnetic transition in the field range of investigation. The magnetization varies linearly up to the field value of 7 T for the compound GdTi$_2$Al$_{20}$ and saturates to a value of 6.16  $\mu_{B}$/fu at 1.9 K whereas the vanadium sample saturates to the value of 5.2  $\mu_{B}$/fu (Refer Fig.7c, 7d and Table3). Assuming localised magnetic character, the deficiency in saturation magnetization and effective magnetic moment can be explained by the following arguments. It has been observed in 1-2-20 family of compounds that the transition temperature varies linearly with different rare-earth and exchange parameter scales with the de Gennes factor G = $(g_J-1)^2(J(J+1))$ \cite{burnett}. Hence the governing interaction in these systems is conduction electron mediated indirect exchange interaction. In order to understand the induced spin-polarization and short range correlations it is important to compare the relative intensity of various exchange parameters viz., $J_{4f-sp}$; $J_{4f-3d}$; $J_{4f-5d}$. In CeCr$_2$Al$_{20}$ structure, R-R; T-T and R-T distances are quite large and R atoms are surrounded exclusively by Al atoms. Hence the negative induced polarization (local) due to the $J_{4f-sp}$ interactions would be stronger than other intra-atomic and inter-atomic exchange interactions. X-ray magnetic circular dichroism measurements in the presence of 0.8 T on GdFe$_2$Zn$_{20}$ by Mardegan et al showed evidently an induced magnetic signal of 0.06 \% at the Zn-K edge \cite{XRMS}. Further, they have observed magnetic intensities of Gd and Zn ions following the same temperature variation below 85 K and vanishing in the paramagnetic regime suggesting spin polarization of Zn \textit{4p} states. Another interaction which may be relevant in this scenario is the intra-atomic exchange between the localised 4f states and 5d bands $J_{4f-5d}$ suggested by Campbell in Laves phase compounds \cite{campbell} \cite{burzo}. He proposed an empirical model that in cubic Laves phase compounds the $J_{4f-5d}$ intra-atomic exchange induces 5d band polarization and positive local d moment \cite{campbell}. This model explained the observed deviation from the saturated magnetization in Laves phase compounds and Gd alloys \cite{campbell}. The CeCr$_2$Al$_{20}$ type structure is closely related to the ZrZn$_{22}$ system which is derived from the RTM$_2$ (Laves phase) unit cell. The inter-atomic distances (R-R) of cubic Laves phase and CeCr$_2$Zn$_{20}$ structure are $\approx$ 5.2 $\textrm{\AA}$  and $\approx$ 6.0 $\textrm{\AA}$ respectively \cite{svanidze}. Hence the $J_{4f-5d}$ intra-atomic exchange could also play an important role in the reduced magnetic moment of the title compounds. However in order to claim the presence of 5d band polarization and  $J_{4f-5d}$ intra-atomic exchange interaction band structure and density of states calculations are needed.
\subsection{Heat Capacity} 
The specific heat of a solid in the absence of low frequency optical lattice vibration modes can be written of the form
\\ 
$C_{\small P}(T)$ = $\gamma T$+ $C_{ph,D} (T)$ \qquad \qquad \textrm{(6)}
\\
$C_{ph,D} (T)$ = $\Big[9NR(\frac{T}{\Theta_D})^3\int\limits_0^\frac{\Theta_D}{T}\frac{x^4}{(e^x-1)(1-e^{-x})}dx \Big] \qquad \qquad \textrm{(7)}$
\\

The first and second terms in equation (6) represent the electronic specific heat and Debye lattice heat capacity respectively. In equation (7),  $\Theta_D$ stands for Debye temperature and $N$ represents the total number of atoms in the formula unit \cite{ESR}. The above expression accounts for acoustic phonon modes which is characterized by the Debye temperature $\Theta_D$. 
The Einstein specific heat accounts for optical mode of vibrations and can be expressed as follows
\\
$C_{ph,E} (T) =\sum_{i}^{}n_{Ei}R(\frac{\Theta_{Ei}}{T}) \frac{e^{\frac{\Theta_{Ei}}{T}}}{\big(e^{\frac{\Theta_{Ei}}{T}} -1\big)^2} \qquad \qquad \textrm{(8)}$
\\
$\Theta_{Ei}$ represents Einstein temperature of  $i^{th}$ group with $n_{Ei}$ optical modes. We have attempted to iteratively fit the experimental heat capacity data by combining both  $C_{ph,D} (T)$ and   $C_{ph,E} (T)$. The fit using 3 Einstein oscillators corresponding to the three atoms (R and two TM atoms)  along with Debye's acoustic contribution did not converge to reasonable values. However the presence of low frequency optical modes can be qualitatively analyzed using the $C_p(T)/T^3$ vs T plot. Fig.8 shows a maximum at ($T_{max}$) $\sim$ 27 K for all the compounds plotting $C_P(T)/T^3$ against temperature . The peak positions for the non-magnetic compounds agree well with the reported values\cite{Sakai}. The blue solid line in figure 9 represents a simulation of the Einstein heat capacity data from the equation (8).  The simulation is done with the  assumption that 3 atoms per formula unit exhibit optical modes of vibration and with Einstein temperature $\Theta_{E}$ = 140 K ($T_{max}$ =  $\Theta_{E}/5$). The red line represents Debye heat capacity calculated using the equation (7) by assuming $\Theta_D$ = 480 K. The peak position ($T_{max}$) moderately scales with molecular weight observed from EPMA measurements. For example, the lowest and highest peak positions are observed at 22 K for GdV$_2$Al$_{20}$ and 32 K for EuV$_2$Al$_{20}$ compounds which correspond to the respective low and high value of the concentration ratio between Al and R indicating the amplitude and frequency of Einstein vibrational modes that depend upon the molecular weight, rare-earth vacancy and inter-atomic distance (d$_{R-Al1}$). 
\begin{figure}[h!]
	\centering
	\includegraphics[scale=0.35]{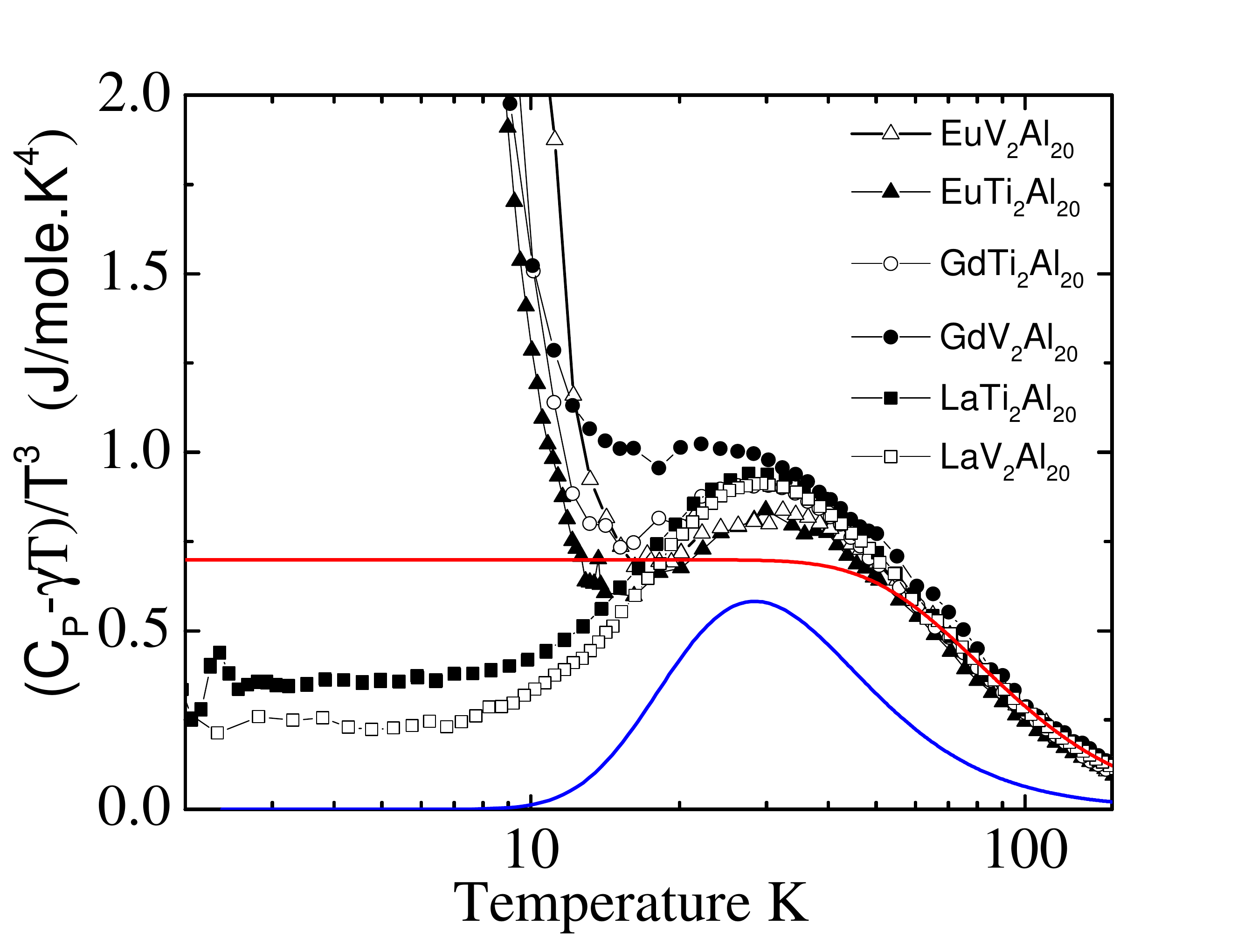}
	\caption{(Color online) The temperature variation of the function ($C_P- \gamma T/T^3$)$^*$ plotted to indicate the presence of low frequency Einstein Modes. The red solid line represents the simulated Debye's heat capacity divided by T$^3$ assuming $\Theta_D$ = 480 K. The blue solid line is the temperature variation $C_{ph,E}/T^3$estimated using the equation(8)
		\newline
		$^*$ \textit{The electronic contribution is subtracted only for the non-magnetic compounds.}
	}
	\label{Fig1}
\end{figure}
The Sommerfeld coefficient for the non magnetic compound is estimated according to the relation 
\\
$C_P(T)/T$ = $\gamma$ + $\beta T^2$;  $\gamma=(2/3)\pi^2k_B^2N(E_F)\qquad \qquad \textrm{(9)}$.
\\ 
As seen in the figure 9 the lattice contribution is dominant even in the low temperatures and found not to follow the Debye $T^3$ law.  Hence the Sommerfeld coefficient is extracted only for the non magnetic compounds by fitting the experimental data $<$ 10 K. However comparison of $C_p(T)/T$ vs $T^2$ variation between the  compounds indicate that the Sommerfeld coefficient values of the magnetic ion bearing compounds are of the same order as in the La(Ti/V)$_2$Al$_{20}$ compounds (See Fig.9a and 9b) except for the compound GdV$_2$Al$_{20}$ in which the heat capacity coefficient is slightly larger compared to the other compounds. It is interesting to note that addition of one electron between V and Ti derivatives alter the band structure and density of states drastically in the case of Gd compounds similar to the observation noticed by Jia et al. in GdFe$_2$Zn$_{20}$ and GdCo$_2$Zn$_{20}$. They proposed through band structure calculation that addition of two electrons in GdCo$_2$Zn$_{20}$  manifest higher electronic density of states at the Fermi level compared to the $N(E_F$) values for GdFe$_2$Zn$_{20}$ \cite{Canfield3}.

\begin{figure}[htb!]
	\centering
	\includegraphics[scale=0.33]{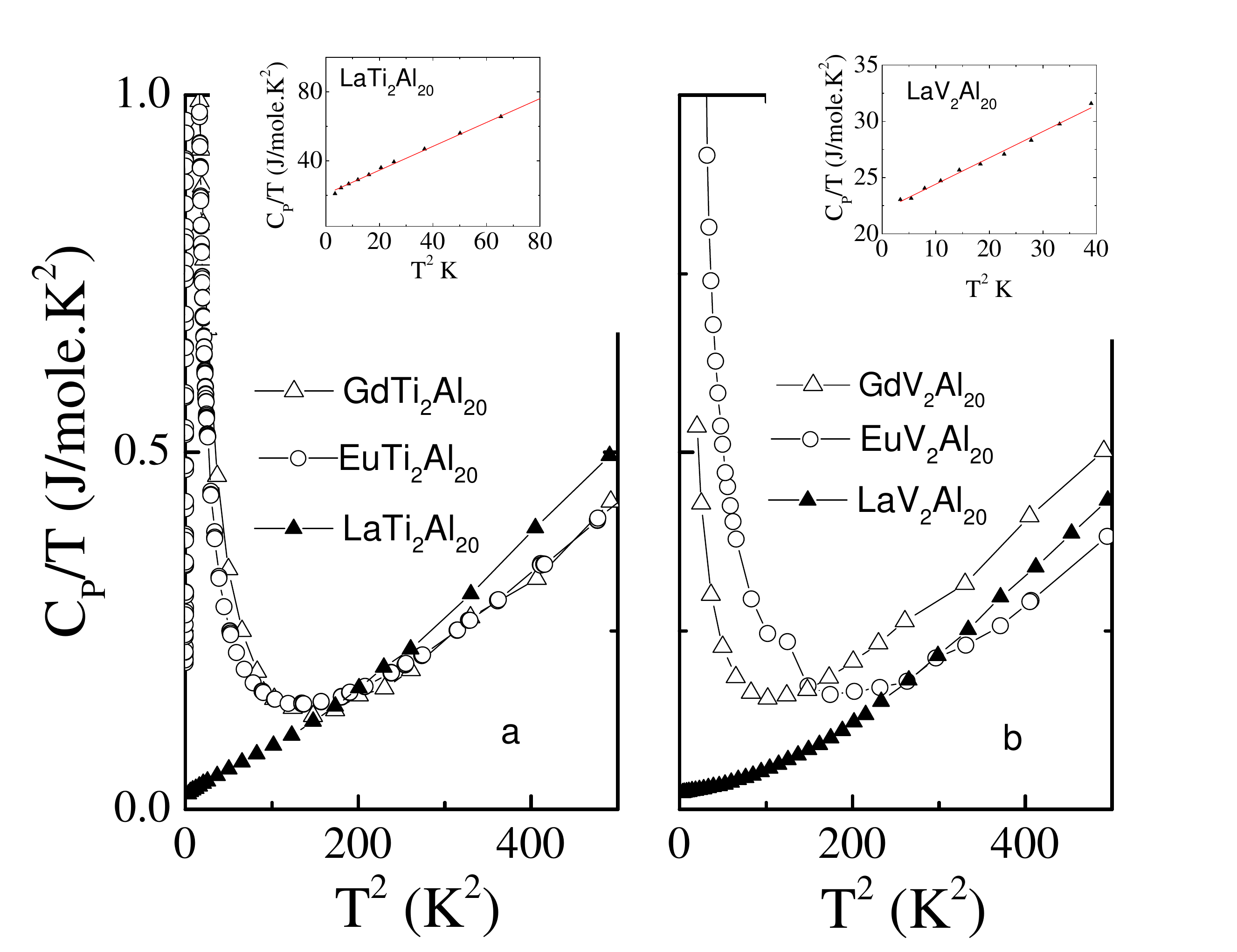}
	\caption{(Color online)Squared temperature variation of ($C_P/T$) for the compounds RTM$_2$Al$_{20}$ (R = Eu, Gd and La; TM = V and Ti). Red solid line in the inset picture represents  linear fits to the low temperature data of the two non-magnetic La compounds}
	\label{Fig1}
\end{figure}
In the low temperatures, (Gd/Eu)Ti$_2$Al$_{20}$ showed markedly different heat capacity variation compared to the vanadium counterparts see Fig.10a and 10c. The heat capacity of the Ti compounds showed a huge and abrupt jump at the onset of magnetic ordering viz., 89.1 J/mol.K in EuTi$_2$Al$_{20}$ and 61.1 J/mol.K in the case of GdTi$_2$Al$_{20}$. 
\begin{figure}[h!]
	\centering
	\includegraphics[scale=0.36]{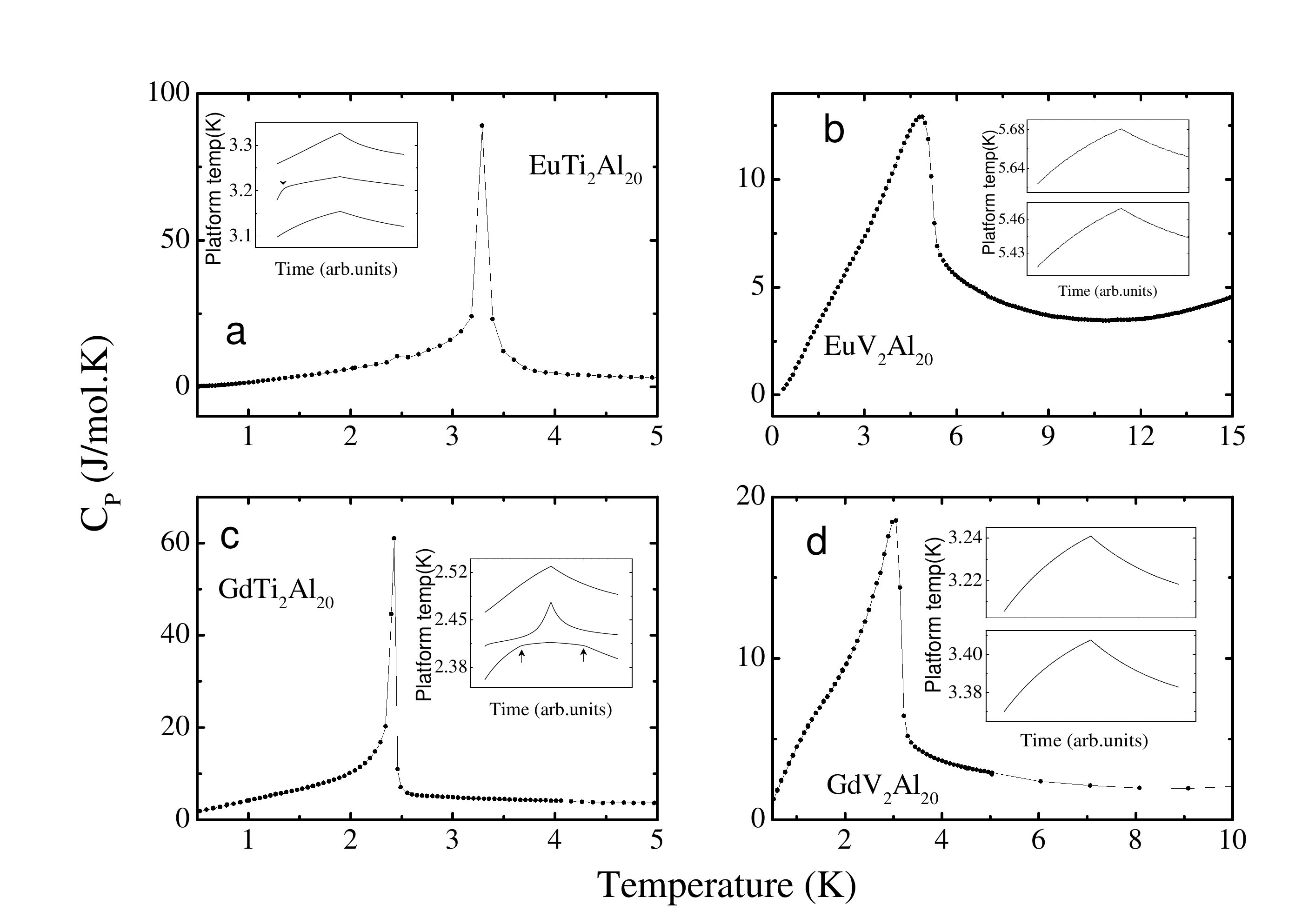}
	\caption{Low temperature specific heat of RTM$_2$Al$_{20}$ (R = Eu, Gd ; TM = V and Ti) plotted as $C_P$ versus T. The inset plots in all the main panels depict the corresponding temperature-time relaxation curve. The thermal arrests due to latent heat across the first order transition are indicated by arrows.} 
\end{figure}
This behavior is a manifestation of first-order magnetic phase transition which is in concert with the  susceptibility measurements. PPMS employs thermal relaxation method to determine the sample's specific heat by measuring the thermal response of sample and calorimeter assembly for a known change in heating condition through external power P. The detailed discussion about the approximations and conditions under which the sample heat capacity is measured can be found in reference \cite{FOPT}. The Bachmann approach of measuring heat capacity is valid only under the following conditions (i) the time constants of the heat capacity measurements should not be more than 100 s and  (ii) the fractional value should be ($\frac{\Delta T}{C}$) ($\frac{dC}{dT}$) $<<$ 1 \cite{Bach}. The above inequality is not valid for the phase transition which show huge and abrupt jump as observed in the case of Sm$_2$IrAl$_8$ and Gd$_5$Si$_2$Ge$_2$ \cite{SmIr,Pecharsky,GeibelFO}. Our heat capacity measurements showed similar behaviour in the heat capacity. Hence conventional C$_P$ measurements around first-order transition is somewhat compromised in relative accuracy in the vicinity close to the peak temperature, the observed value may not be completely accurate and in fact it becomes dependent upon the measurement conditions. The detailed discussion of C$_P$ across the first-order transition in the compounds Gd/EuTi$_2$Al$_{20}$ will be reported elsewhere. Here we emphasize that the relaxation curve across the transition showed slope change in the heating cycle for the compound EuTi$_2$Al$_{20}$ whereas both heating and cooling cycles showed abrupt change for GdTi$_2$Al$_{20}$ (Refer inset of figure 10). The slope change in the relaxation curves is a direct evidence for thermal arrest and presence of latent heat which indicates the magnetic transition in (Gd/Eu)Ti$_2$Al$_{20}$ belong to first-order type transition. By contrast, vanadium samples (Gd/Eu)V$_2$Al$_{20}$ showed a linear variation of C$_P$ and their corresponding relaxation curves resemble a typical behaviour for a second order transitions.  
Within the framework of mean field theory (MFT), the specific heat jump at the onset of a continuous (non-first order) equal moment - antiferromagnetic (em-afm) state is given by
\\
$\Delta C$ (T = T$_{mag}$) =  5R $(\frac{J(J+1)}{2J^2+2J+1}). \qquad \qquad \textrm{(10)}$
\\
and by assuming Eu (2+) and Gd(3+) electronic state the specific heat jump for a second order transition is equal to 20.14 J/mol.K \cite{salamakah}. The heat capacity jump for the sample EuV$_2$Al$_{20}$ at the transition point was observed to be 12.9 J/mol.K whereas the GdV$_2$Al$_{20}$ sample showed a jump of 18.53 J/mol.K. Considering the deficiency of rare-earth ions, the heat capacity jump in GdV$_2$Al$_{20}$ signals a collinear spin arrangement as observed in iso-structural compounds viz GdFe$_2$Al$_{20}$ and GdCo$_2$Al$_{20}$. In contrast to the Gd compound we observed 36 $\%$ reduction in the heat capacity jump for the compound EuV$_2$Al$_{20}$. The reason for this behaviour is not clear as there are several reasons which can lead to a reduced $C_P$ jump. Firstly, even though the temperature variation of $C_P$ showed a $\lambda$ type transition an abrupt jump is evident close to $T_N$ and hence the observed peak value strongly depends upon the density of points of the $C_P$ measurements collected at the apex of the peak. Secondly, the broad hump in the magnetic susceptibility suggest a presence of either FM transition or spin re-orientation type transition below $T_N$. We believe that the reduced value of $C_P$ could be due to a complex spin arrangement which cannot be accounted by simple collinear equal moment-AFM or amplitude modulated-AFM ordering.     
We estimated the magnetic heat capacity $C_{mag}(T)$ by subtracting the heat capacity of non magnetic compounds from the experimental heat capacity of RTM$_2$Al$_{20}$ (R = Eu, Gd; TM = V and Ti) compounds. The magnetic entropy $S_{mag}(T)$ was calculated from the following expression. 
\\
$S_{mag}(T)$ = $\int\limits_0^T\frac{C_{mag}}{T'}dT'$ \qquad \qquad \textrm{(11)}.
\\ 
The distinction observed between the vanadium and titanium derivatives in the $C_P(T)$ is seen in the entropy estimation as well (Fig.11). The magnetic entropy $S_{mag}(T)$ shows a huge jump near the ordering temperature for the   (Gd/Eu)Ti$_2$Al$_{20}$ compounds whereas temperature variation of vanadium compounds exhibit linear increase in the ordered state. From figure 11, it is observed that the $S_{mag}(T)$ ($T_N$ $<$ T  $\rightarrow$ $\infty$) is saturating to the value somewhat less than $R\ln8$ expected for pure spin systems (Table 4). 
\begin{figure}[h!]
	\centering
	\includegraphics[scale=0.35]{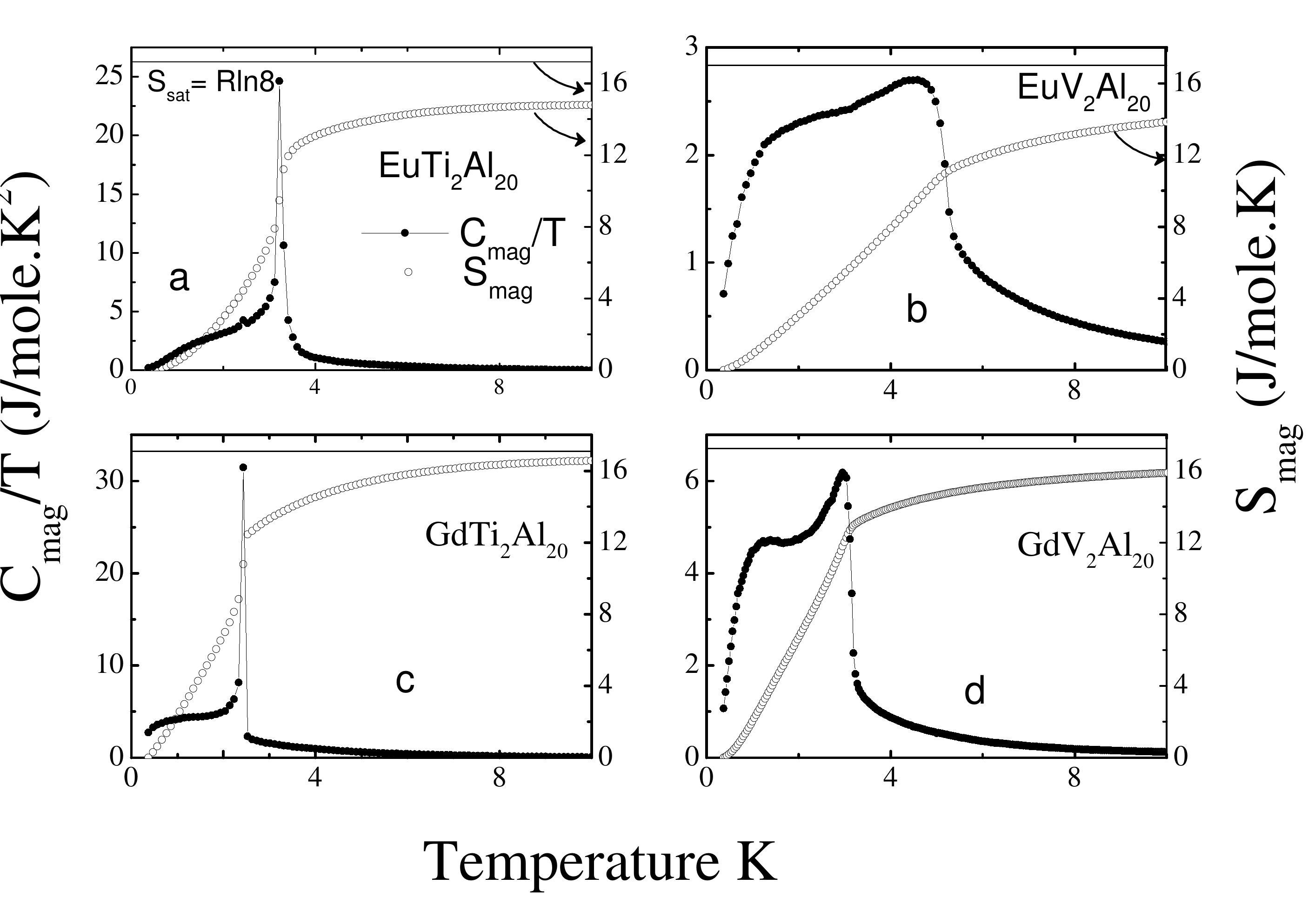}
	\newline
	\caption{(a-d) Magnetic heat capacity $C_{mag}(T)$ and magnetic entropy $S_{mag}$ for the compounds RTM$_2$Al$_{20}$ (R = Eu, Gd ; TM = V and Ti) plotted versus temperature. The magnetic contribution is estimated by subtracting the heat capacity of corresponding non-magnetic compound La(Ti/V)$_2$Al$_{20}$. The left and right scales correspond to $C_{mag}(T)$ and $S_{mag}$ respectively.}
\end{figure}
The compound GdTi$_2$Al$_{20}$  recovered 95 \%\ of the magnetic entropy in the high temperature limit ($T_N$ $<$ T  $\rightarrow$ $\infty$). The other compounds exhibits lesser values expected for the pure spin system and the reason for lower saturation values could be due to the possible deficiency of rare-earth stoichiometry or error in the accurate estimation of phonon contribution. Finally, within MFT the temperature variation $C_{mag}(T)/T$ is expected to vary as $T^3$ for long range 3D AFM \cite{VKAnand}. We observed a clear anomaly in  $C_{mag}(T)/T$ below the ordering temperature for all the compounds. The T-linear behaviour followed by a broad hump in $C_{mag}(T)/T$ near $T_N/3$ signals possible FM ordering inside of the AFM state as explained by Fishman and Liu \cite{Fishman}. In general, the plateau like behaviour is observed in $C_{mag}(T)/T$ for high S values due to the splitting of ground state multiplet by the internal magnetic field below the magnetic ordering \cite{Sullow}. Our physical property measurement and detailed analysis hints at a double magnetic transition but without microscopic probes it may not be possible to  conclude the reason for the broad hump in $C_{mag}(T)/T$ below $T_N$.
\begin{center}
	\begin{table*}[!t]
	\caption{\label{tab1} Summary of physical properties for the compounds RTM$_2$Al$_{20}$ (R = Eu, Gd and La; TM = V and Ti).  Magnetic entropy  $S_{mag}^{sat}$, electron-phonon coupling constant B, \textit{s-d} inter-band scattering coefficient $\alpha^R$, and residual resistivity ($\rho_0$).}
	\setlength{\tabcolsep}{9pt}
	\begin{tabular}{c c c c c c c c}
		\hline
		\hline
		& & & & & & &   \\ 
		
		& Compound  & $T_{max}$ & $S_{mag}^{sat}$ & $B$ & $\Theta_R$ & $\rho_0$ & $\alpha^R$ \\
		&   & (K) & mJ/mol.K  & $ \mu\Omega.\textrm{cm.K} $ & $(\textrm{K})$ &  $ \mu\Omega. \textrm{cm} $ & $\times10^{-13}\Omega.\textrm{cm.} \textrm{K}^{-3}$  \\

		\hline
		& & & & & & &   \\ 
		& EuTi$_2$Al$_{20}$   & 29 & 14.7 & .0843 & 276 & 86.1 & 2.9  \\
		& & & & & & &  \\
		& EuV$_2$Al$_{20}$   & 32 & 13.7 & 0.073 & 292 & 89.4 & 19.5 \\
		& & & & & &  &  \\
		& GdTi$_2$Al$_{20}$  & 28 & 16.4 & 0.129 & 324 & 14.4 & 6.42 \\
		& & & & & & &  \\
		& GdV$_2$Al$_{20}$  & 22 & 15.8 & 7.04 & 266 & 24.2 & 2.71 \\
		& & & & & & &  \\
		& LaTi$_2$Al$_{20}$  & 27 & - & - & - & - & - \\
		& & & & & & &  \\
		& LaV$_2$Al$_{20}$  & 28 & -  & - & - & - & - \\
		& & & & & & & \\
		\hline
	\end{tabular}
\end{table*}
\end{center}
\subsection{Resistivity}
In general, the resistivity of rare-earth intermetallic compounds can be written as
\newline
$\rho$ = $\rho_0$+$\rho_{e-e}(T)$+$\rho_{el-ph}(T)$+$\rho_{mag}(T).  \qquad \qquad \textrm{(12)}$
\newline
$\rho_0$ is thetemperature independent residual resistivity which depends collectively upon crystal lattice defects such as dislocations, $\rho_{e-e}(T)$ is the resistivity due the electron-electron interaction which is prominent only in the low temperatures, and the $\rho_{el-ph}(T)$ term is temperature dependent electrical resistivity due to the scattering of conduction electrons by lattice vibrations (phonons). In the ordered state the resistivity is having an additional contribution due to the scattering of charge carriers by a magnetic term $\rho_{mag}$ which can have various contributions such as Fermi-liquid quasiparticle excitations represented by parabolic temperature variation and spin-wave magnon excitations with characteristic energy gap $\Delta$ \cite{Mott1}.
\\
Figure 12 shows the temperature variation of electrical resistivity for the compounds RTM$_2$Al$_{20}$ (R = Eu, Gd  TM = V and Ti) in the temperature range between 1.8 K to 300 K. In high temperatures, the resistivity is expected to show a linear dependence which is a consequence of the T $\rightarrow$ $\infty$ limit expansion of the Debye integral formulation of $\rho(T)$ and the domination of simple electron-phonon scattering.   
\begin{figure}[h]
	\centering
	\includegraphics[scale=0.36]{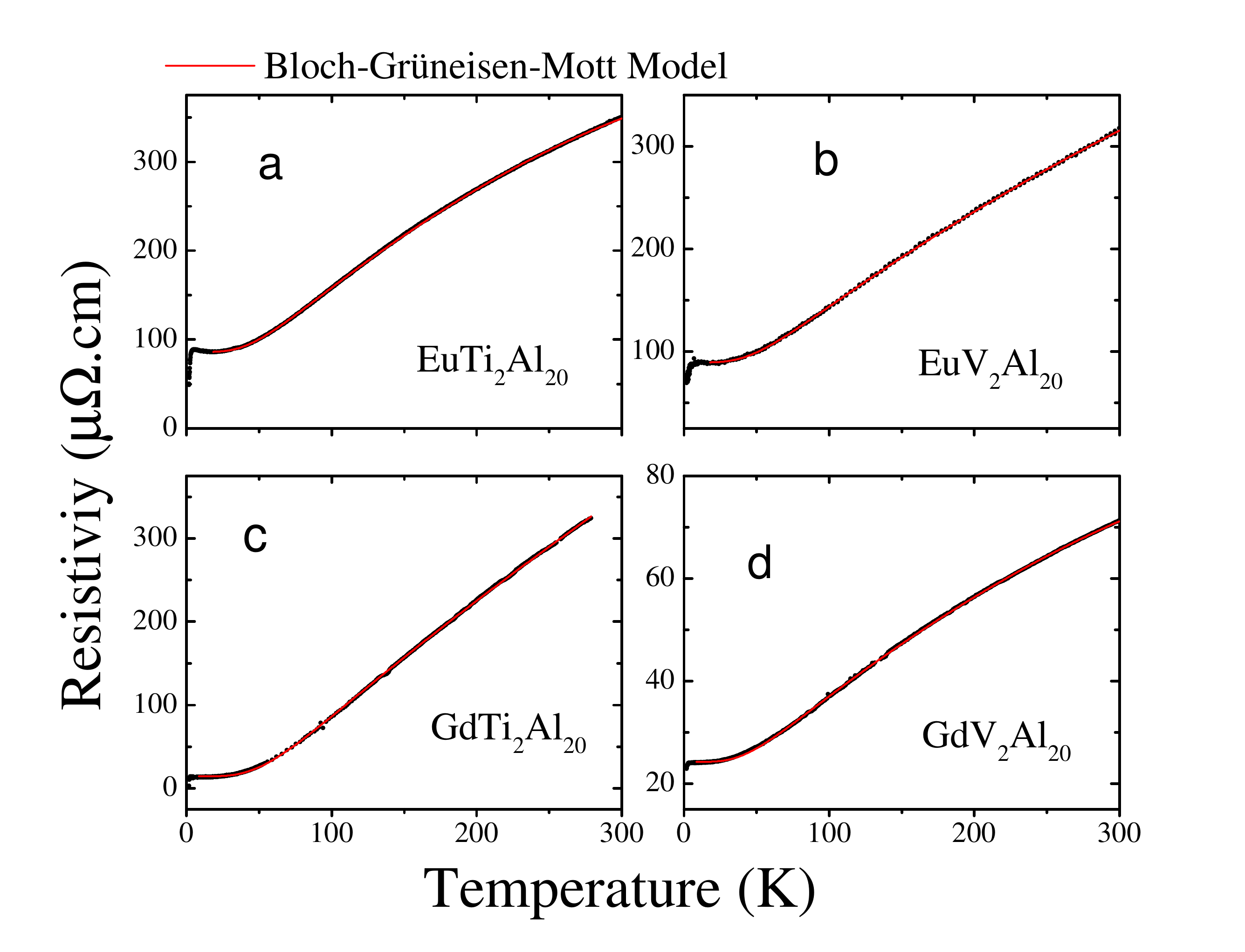}
	\newline
	\caption{Temperature variation of zero field electrical resistivity between 1.8 K and 300 K. In the paramagnetic state the resistivity is fitted with Bloch-Gr$\ddot{\textrm{u}}$neisen-Mott relation (see equation13).} \label{Fig1}
\end{figure}
The electrical resistivity of the conduction electrons due to the scattering by the acoustic phonons can be written as 
\newline
$\rho(T)$=$\rho_0$+$\frac{4B}{\Theta_R}(\frac{T}{\Theta_R})^5$$\int\limits_0^\frac{\Theta_R}{T}\frac{x^5}{(e^x-1)(1-e^{-x})}dx$-$\alpha T^3  \qquad \qquad \textrm{(13)}$

where B and $\Theta_R$ are the electron-phonon coupling parameter and Debye temperature respectively \cite{Mott1,pade}. The term $\alpha T^3$ is added to the resistivity expression to account for s-d interband scattering wherein the parameter $\alpha$ is known as Wilson coefficient. Figure 12 shows that the BGM equation fits satisfactorily above the N{\'e}el temperature. As a representative example, the fitting parameters for the compound EuTi$_2$Al$_{20}$ are B = 0.084 $\mu\Omega$cm.K$\Theta_R$ = 276 K, $\rho_0$ = 86 $\mu\Omega$.cm; $\alpha$ = 2.9 $\times10^{-13}\Omega.\textrm{cm.} \textrm{K}^{-3}$ . The fitting values for the compounds based on BGM model are given in Table 2. In the ordered state, the temperature variation of resistivity shows a sudden drop due to the loss of spin disorder contribution. The compounds (Gd/Eu)Ti$_2$Al$_{20}$ exhibit a discontinuous jump at 3.2 K and 2.2 K respectively below the $T_N$ in concert with magnetization and heat capacity measurements. The first-order nature of the phase transition in the compounds (Gd/Eu)Ti$_2$Al$_{20}$ is clearly seen in temperature variation of $d\rho/dT$ plot (Figure13 panel a and c)
\begin{figure}[h!]
	\centering
	\includegraphics[scale=0.36]{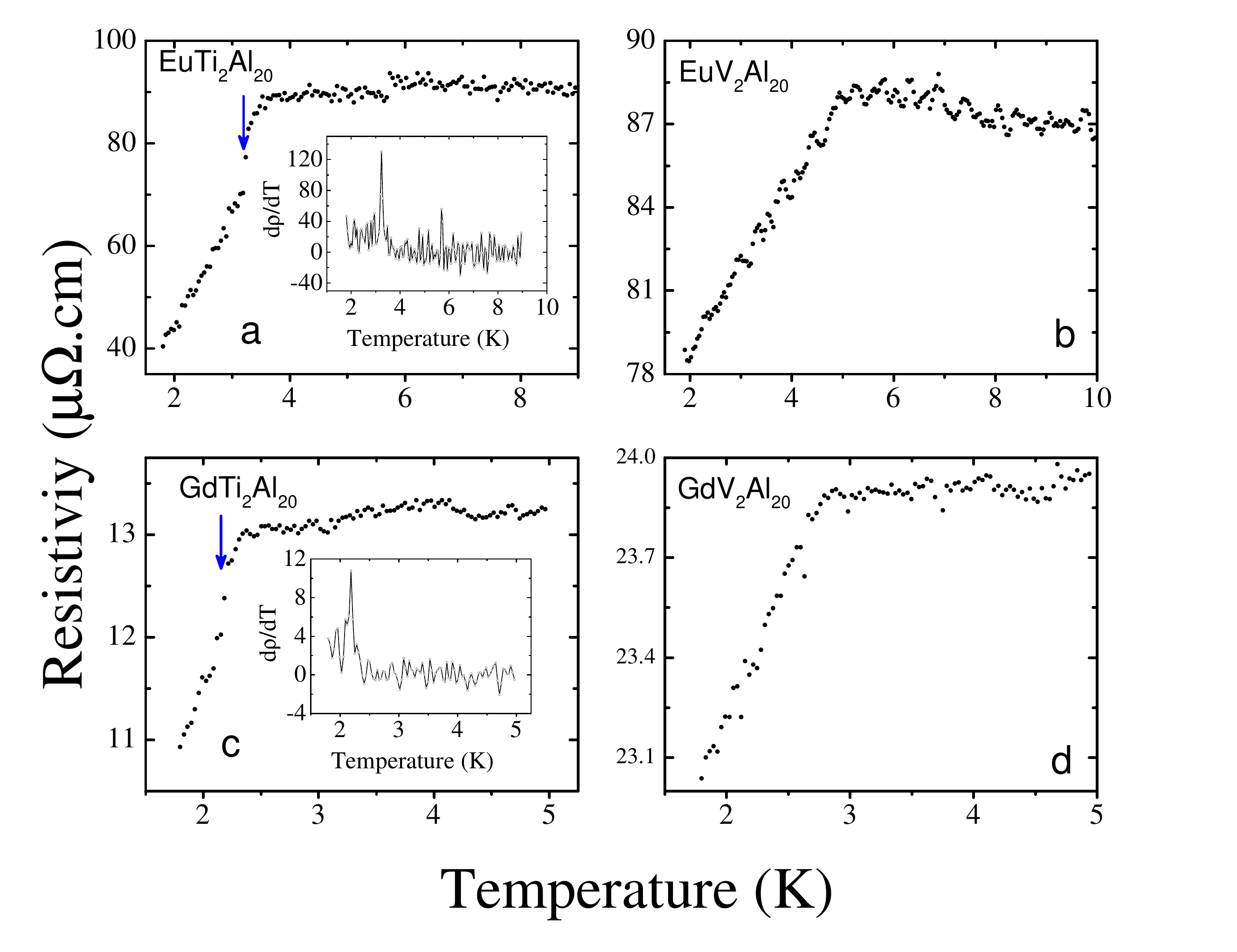}
	\caption{Temperature variation of electrical resistivity close to the magnetic ordered state. Arrows indicate the first-order type magnetic transition in (Gd/Eu)Ti$_2$Al$_{20}$} 
\end{figure}

\section{Summary and Conclusions}

The compounds RTM$_2$Al$_{20}$ (R = Eu, Gd and La; TM = V and Ti) form in cubic CeCr$_2$Al$_{20}$ type crystal structure and their physical properties were investigated through magnetic, electrical and thermal transport measurements.  Rietveld analysis and EPMA measurements indicated 2-10 $\%$ of rare-earth ion deficiency in various compounds. The lattice parameter variation between different rare-earth elements showed an anomaly for the Eu based compounds indicating divalent electronic state of Eu ions in this structure. The magnetic susceptibility $\chi(T)$, heat capacity $C_P(T)$ and resistivity $\rho(T)$ measurements show long range anti-ferromagnetic ordering in the temperature range of 2.4 to 5.6 K. The effective moment showed notable deficiency from the full moment value expected for the pure spin system S = 7/2. The compound EuV$_2$Al$_{20}$ showed lowest effective moment value of 6.36 $\mu_B$ which accounts for 20 percent deficiency from the theoretical value ($\mu_{eff}$ = 7.94 $\mu_B$). Similar behavior is observed for the other magnetic systems as well despite observing nominal stoichiometry with 98 \% rare-earth occupation rate. In concert with the susceptibility measurements, $M(H)$ along $<111>$ in the ordered state (T $< T_N$) saturated to much lower values expected for local moment systems. We propose that the induced negative polarization by 4f local moment reduces the on-site effective moment which in turn decreases the bulk magnetization in these compounds. The CW behaviour free V-ions and 3d local moment due to $J_{4f-5d}$ intra-atomic exchange interaction could also play a role in the observed low magnetic moment in EuV$_2$Al$_{20}$ and GdV$_2$Al$_{20}$. Magnetization studies showed three successive spin-flop meta magnetic transitions for the compound EuTi$_2$Al$_{20}$. The hysteresis behaviour between increasing and decreasing fields in EuTi$_2$Al$_{20}$ indicate the first-order nature of the phase transition and in contrast the metamagnetism observed at 1 T for  EuV$_2$Al$_{20}$ seemingly a continuous phase transition. Large diamagnetic response is seen for the compound LaV$_2$Al$_{20}$ whereas the LaTi$_2$Al$_{20}$ sample showed temperature independent Pauli paramagnetism. The low temperature susceptibility measurements and analysis show an addition orbital contribution from Landau-Peierls term responsible for the large negative susceptibility for the case of LaV$_2$Al$_{20}$. Abrupt and large value of $C_P$ in the vicinity of magnetic transition in (Gd/Eu)Ti$_2$Al$_{20}$ indicating a possible discontinuous transition and the slope change in the temperature-time relaxation plot is taken as direct evidence of first-order nature. The $C_P/T^3$ vs T data showed the presence of low frequency Einstein modes of vibrations. The peak position in the  $C_P/T^3$ vs T plot is successfully simulated using the Einstein heat capacity expression by assuming 3 Einstein modes that consistently correspond to three guest ions occupying the oversized polyhedron in the crystal structure.  The molar magnetic entropy follows a typical temperature variation for pure spin systems with S=7/2. Resistivity measurements showed  a metallic temperature variation the data could be modeled using a standard electron-phonon scattering formalism. The resistivity decreases rapidly on cooling below $T_N$ due to loss of spin-disorder scattering and sudden jump in the resistivity for the Ti compounds confirms the first-order magnetic transition.

\begin{center} 
	\bf Acknowledgments
\end{center}
The authors, RKK and HSN acknowledge the FRC/URC Postdoctoral fellowship. AMS thanks the SA-NRF (93549)  and the FRC/URC of UJ for financial assistance. RKK thanks Ruta Kulkarni, scientific officer,  DCMP\&MS, TIFR and Michael O. Ogunbunmi, UJ for their help in carrying out the low temperature resistivity  measurements.
%
\\

{\bf References}
\newline
\end{document}